\documentclass[12pt]{article} 

\usepackage{epsfig}
\usepackage{graphicx}
\usepackage{comment}
\usepackage{latexsym}
\usepackage{hyperref}
\usepackage{amsmath}
\usepackage{color}
\usepackage{amsbsy}
\usepackage{amssymb}
\usepackage{amsthm}
\usepackage{amsfonts}
\usepackage{cite}

\newcommand{\be}[0]{\begin{equation}}
\newcommand{\ee}[0]{\end{equation}}

\newcommand{\F}{{\cal F}}
\newcommand{\I}{{\cal I}}
\newcommand{\J}{{\cal J}}
\newcommand{\N}{{\cal N}}
\newcommand{\V}{{\cal V}}
\renewcommand{\S}{{\cal S}}
\newcommand{\A}{{\cal A}}
\newcommand{\G}{{\cal G}}
\renewcommand{\H}{{\cal H}}

\newcommand{\Z}{\mathbb{Z}}

\newcommand{\Ka}{K{\"a}hler }
\renewcommand{\O}{{\cal O}}
\renewcommand{\Re}{{\rm Re}\,}
\renewcommand{\Im}{{\rm Im}\,}

\newcommand{\abs}{|}

\newcommand{\Str}{\textrm{Str}\,}
\newcommand{\rk}{\textrm{rk}}

\newcommand{\ie}{{\em i.e.} }
\newcommand{\eg}{{\em e.g.} }
\newcommand{\via}{{\it via} }

\newcommand{\where}{\mbox{where}}
\newcommand{\with}{\mbox{with}}

\renewcommand{\and}{\mbox{and}}

\newcommand{\desp}{\!\!\!\phantom{\Bigg\abs}}
\newcommand{\tesp}{\!\!\! \phantom{\underset{\hat \abs}{\Big\abs}}}

\newcommand{\bm}{\boldmath} 

\catcode`\@=11
\def\marginnote#1{}
\newcount\hour
\newcount\minute
\newtoks\amorpm
\hour=\time\divide\hour by60 \minute=\time{\multiply\hour by60
\global\advance\minute by-\hour}
\edef\standardtime{{\ifnum\hour<12 \global\amorpm={am}%
        \else\global\amorpm={pm}\advance\hour by-12 \fi
        \ifnum\hour=0 \hour=12 \fi
        \number\hour:\ifnum\minute<10 0\fi\number\minute\the\amorpm}}
\edef\militarytime{\number\hour:\ifnum\minute<10 0\fi\number\minute}
\def\draftlabel#1{{\@bsphack\if@filesw {\let\thepage\relax
   \xdef\@gtempa{\write\@auxout{\string
      \newlabel{#1}{{\@currentlabel}{\thepage}}}}}\@gtempa
   \if@nobreak \ifvmode\nobreak\fi\fi\fi\@esphack}
        \gdef\@eqnlabel{#1}}
\def\@eqnlabel{}
\def\@vacuum{}
\def\draftmarginnote#1{\marginpar{\raggedright\scriptsize\tt#1}}
\def\draft{\oddsidemargin -.2truein
        \def\@oddfoot{\sl preliminary draft \hfil
        \rm\thepage\hfil\sl\today\quad\militarytime}
        \let\@evenfoot\@oddfoot \overfullrule 3pt
        \let\label=\draftlabel
        \let\marginnote=\draftmarginnote
   \def\@eqnnum{(\theequation)\rlap{\kern\marginparsep\tt\@eqnlabel}%
\global\let\@eqnlabel\@vacuum}  }
\def\thebibliography#1{
\vskip 0.5cm \centerline{\bf References}
\list{
[\arabic{enumi}]}{\settowidth\labelwidth{[#1]}
\leftmargin\labelwidth
\advance\leftmargin\labelsep
\usecounter{enumi}}
\def\newblock{\hskip .11em plus .33em minus .07em}
\sloppy\clubpenalty4000\widowpenalty4000
\sfcode`\.=1000\relax}

\renewcommand{\theequation}{\arabic{section}.\arabic{equation}}
\renewcommand{\section}{\setcounter{equation}{0}\@startsection
{section}{1}{0mm}{-\baselineskip}{0.5\baselineskip} {\normalfont\Large\bfseries}}
\renewcommand{\subsection}{\@startsection
{subsection}{2}{0mm}{-\baselineskip}{0.5\baselineskip} {\normalfont\large\bfseries}}
\renewcommand{\subsubsection}{\@startsection
{subsubsection}{3}{0mm}{-\baselineskip}{0.5\baselineskip}
{\normalfont\normalsize\slshape}}

\begin{document}


\begin{titlepage}
\begin{flushright}
LPTENS--16/04,
CPHT--RR033.062016,~
July   2016
\vspace{-.0cm}
\end{flushright}
\begin{centering}
{\bf \Large Super no-scale models in string theory}

\vspace{5mm}

 {\bf Costas Kounnas$^{1}$ and Herv\'e Partouche$^2$}

 \vspace{1mm}

$^1$ Laboratoire de Physique Th\'eorique,
Ecole Normale Sup\'erieure$^\dag$,\\
24 rue Lhomond, F--75231 Paris cedex 05, France\\
{\em  Costas.Kounnas@lpt.ens.fr}

$^2$  {Centre de Physique Th\'eorique, Ecole Polytechnique, CNRS, Universit\'e Paris-Saclay\\
F--91128 Palaiseau cedex, France\\
{\em herve.partouche@polytechnique.edu}}

\end{centering}
\vspace{0.1cm}
$~$\\
\centerline{\bf\Large Abstract}\\
\vspace{-0.6cm}

\begin{quote}
We consider ``super no-scale models" in the framework of the heterotic string, where the $\N=4,2,1\to 0$ spontaneous breaking of supersymmetry is induced by geometrical fluxes realizing a stringy Scherk-Schwarz perturbative mechanism. Classically, these backgrounds are characterized by a boson/fermion degeneracy at the massless level, even if supersymmetry is broken. At the 1-loop level, the vacuum energy is exponentially suppressed, provided the supersymmetry breaking scale is small, $m_{3/2}\ll M_{\rm string}$.  We show that the ``super no-scale string models" under consideration are free of Hagedorn-like tachyonic singularities, even when the supersymmetry breaking scale is large, $m_{3/2}\simeq M_{\rm string}$. The vacuum energy decreases  monotonically and converges exponentially to zero, when $m_{3/2}$ varies from $M_{\rm string}$ to $0$. 
We also show that all Wilson lines associated to asymptotically free gauge symmetries are dynamically stabilized by the 1-loop effective potential, while those corresponding to non-asymtotically free gauge groups lead to instabilities and condense. The Wilson lines of the conformal gauge symmetries remain massless. When stable, the stringy super no-scale models admit low energy effective actions, where decoupling gravity yields  theories in flat spacetime, with softly broken  supersymmetry. 

\end{quote}

\vspace{3pt} \vfill \hrule width 6.7cm \vskip.1mm{\small \small \small
  \noindent
   $^\dag$\ Unit{\'e} mixte  du CNRS et de l'Ecole Normale Sup{\'e}rieure associ\'ee \`a l'Universit\'e Pierre et Marie Curie (Paris 6), UMR 8549.}\\

\end{titlepage}
\newpage
\setcounter{footnote}{0}
\renewcommand{\thefootnote}{\arabic{footnote}}
 \setlength{\baselineskip}{.7cm} \setlength{\parskip}{.2cm}

\setcounter{section}{0}


\section{Introduction and summary}

String theory unifies  gravitational and gauge interactions at the quantum level. To describe particle physics, one can naturally consider classical models defined in four-dimensional Minkowski spacetime, where string  perturbation theory  can be implemented to derive the quantum dynamics.  However, from a gravitational point of view, the question of the cosmological constant which can be regenerated at 1-loop, must be addressed. In non-supersymmetric models, such as those derived by compactifying the  $SO(16)\times SO(16)$  ten-dimensional heterotic string, this vacuum energy density is extremely large \cite{GV}. It is generically of order $M_{\rm s}^4$, where $M_{\rm s}$ is the string scale, and has no chance to be naturally cancelled by any mechanism involving physics at lower energy. 

Alternatively,  one can consider no-scale models \cite{noscale}, which by definition describe at tree level theories in Minkowski space, where supersymmetry is spontaneously broken at an arbitrary scale $m_{3/2}$. More precisely, $m_{3/2}$ is a flat direction of a classical positive semi-definite potential, $\V_{\rm tree}\ge0$. This very fact opens the possibility to generate by quantum effects a vacuum energy of arbitrary magnitude. 
In $\N=1$ supergravity language, the no-scale models involve a superpotential $w_0$ and moduli fields $z_i$, in terms of which the scale of the spontaneous supersymmetry breaking can be expressed as\cite{FKZ}, 
\be
m^2_{3/2}= e^{ K}\abs w_0\abs^2= {e^{\tilde K}\abs w_0\abs^2\over \Im z_1\, \Im z_2\, \Im  z_3}\, ,
\ee
where $K$ is the  K\"alher  potential and $\tilde K$ is the part of $K$ that is independent of the three moduli $z_i$ associated to the breaking of supersymmetry.  When $w_0$ is  independent of the $z_i$'s,  $m_{3/2}$ is undetermined by the minimization condition $\langle \V_{\rm tree}\rangle=0$.  In string theory or its associated effective supergravity description at low energy, depending on the choice of supersymmetry breaking mechanism, the $z_i$'s can either be the dilaton-axion field $S$, or \Ka or complex structure moduli $T_I,U_I$ associated to the six-dimensional internal space. For instance :

- Some initially supersymmetric models can develop non-perturbative effects, such as gaugino condensation\cite{gc}. In this case, some of the fields, including $S$, are stabilized. The magnitude of supersymmetry breaking is determined  by $\abs w_0\abs^2={\Lambda_{\rm np}^6/ M_{\rm P}^4}$ and the imaginary parts of $z_i$, $i\in\{1,2,3\}$, which can be  \Ka or complex structure moduli $T_I,U_I$. In the expression of the superpotential, $M_{\rm P}\simeq 2.4\cdot10^{18}$ GeV is  the Planck scale and $\mbox{$\Lambda_{\rm np}=M_{\rm s}\exp{(-8\pi^2 /\abs b\abs g_{\rm s}^2)}$}$ is the  scale of confinement associated to an asymptotically free gauge group, of $\beta$-function coefficient $b$. $g_{\rm s}$ is the string coupling, which relates the string and Planck scales as $M_{\rm s}=g_{\rm s}M_{\rm P}$.  The gaugino condensation breaking mechanism leads naturally to a small gravitino mass, even though the moduli fields $\Im z_i$'s are of order 1. However, this non-perturbative scenario can only be studied qualitatively at the effective  supergravity level, since no fully quantitative derivation from string computations is available yet.

- Alternatively, perturbative or non-perturbative fluxes \cite{Fluxes} along the internal space can induce non-trivial superpotentials that break supersymmetry. In some cases, S-,T- or U-dualities \cite{StringDualities} can be used to derive semi-quantitative results. In general,  there is  not yet available full derivation from string computations  and so, one must restrict to semi-quantitative  descriptions at the effective supergravity level. Some exception however exists, on which we now turn on.

In the present work, we focus on geometrical fluxes that realize 
generalized ``coordinate-dependent compactifications" \cite{SSstring, Kounnas-Rostand}. The latter are similar to that proposed  by Scherk and Schwarz in supergravity \cite{SS},  but upgraded to string theory and furthermore to its gauge sector. In some cases, the mechanism can be implemented at the level of the worldsheet 2$\mbox{-}$dimensional conformal field theory, thus allowing explicit quantitative string computations, order by order in perturbation. The scale $m_{3/2}$ of spontaneous supersymmetry breaking is  given by the inverse volume of the internal directions involved in the generalized stringy Scherk-Schwarz mechanism. For the quantum vacuum energy density to not be of order $M_{\rm s}^4$, this volume should be large, and the associated towers of Kaluza-Klein (KK) states should be light, with many consequences~:

$\bullet$ When their contributions do not cancel each another (a situation that will be central to the present work), the KK states, whose masses are of order $m_{3/2}$, dominate the quantum amplitudes, while the heavier states, whose masses are of order $cM_{\rm s}$, yield exponentially suppressed contributions, $\O(e^{-cM_{\rm s}/m_{3/2}})$. In practice,  $cM_{\rm s}$ can be the string scale, the GUT scale or a large Higgs scale. 

$\bullet$ These dominant contributions are the full expressions obtained in loop computations done in a pure KK field theory that realizes a spontaneous breaking of supersymmetry \`a la Scherk-Schwarz. No UV divergence occurs, a fact that is similar to that observed in field theory at finite temperature when the KK modes are Matsubara excitations along the Euclidean time circle and the spectrum at zero temperature is supersymmetric.  

$\bullet$ At 1-loop, if the model does not contain any scale below $m_{3/2}$, the effective potential takes the form \cite{AntoniadisTeV, ADM,planck2015, N=0thresh}, 
\be
\label{v}
\V_{\mbox{\scriptsize 1-loop}} =\xi(n_{\rm F}-n_{\rm B})\, m_{3/2}^4+\O\!\left(M_{\rm s}^4\, e^{-cM_{\rm s}/ m_{3/2}}\right) ,
\ee 
where $n_{\rm F}$ and $n_{\rm B}$ count the numbers of massless fermionic and bosonic degrees of freedom, while $\xi>0$ depends on moduli fields other than $m_{3/2}$. The above result makes sense in the theories that are free of  ``decompactification problems'' \cite{solving}, which would invalidate the string perturbative approach, due to large threshold corrections  to gauge couplings \cite{thresholds,universality}. For instance,  models realizing either the $\N=4\to 0$ or $\N=4\to 2\to 0$ or $\N=2\to1\to 0$ patterns of  {\em spontaneous} supersymmetry breaking are consistent at the perturbative level \cite{N=0thresh}. 

Notice in Eq. (\ref{v}) the absence of term  proportional to $\Str M^2\, \Lambda^2_{\rm co} \propto m_{3/2}^2 \Lambda^2_{\rm co}$, where $M$ is the mass operator. Such a term appears  in $\N=1$ and $\N=2$ supergravities spontaneously broken to $\N=0$, when the quantum corrections are regularized in the  UV by a cut-off scale $\Lambda^2_{\rm co}=\O(M^2_{\rm s})$.
Even if the extremely large term $m_{3/2}^2 \Lambda^2_{\rm co}$ is not present in string theory, the sub-dominant one, proportional to $m_{3/2}^4$, still occurs  when $n_{\rm F}\neq n_{\rm B}$. This leads a serious difficulty, since it is far too large, compared to the cosmological constant (indirectly) observed by astrophysicists, even when  $m_{3/2}$ is about 10 TeV, which is the order of magnitude of the lowest bound of supersymmetry breaking scale allowed by current observations at the LHC.  

This remark invites us to consider   ``super no-scale models" in string theory \cite{ADM,planck2015}, which are the subclass of no-scale models satisfying the condition $n_{\rm F}=n_{\rm B}$. These theories generate automatically a 1-loop vacuum energy  that is exponentially suppressed, provided  $m_{3/2}$ is much  lower than $cM_{\rm s}$.  The ``super no-scale models'' extend the notion of no-scale structure valid at tree level to the 1-loop level. Note that 
non-supersymmetric classical models satisfying the even stronger property of {\em boson-fermion degeneracy at each mass level} are already know in type II string \cite{L=0, HarveyL=0} and orientifold descendants \cite{1-L=0,Angelantonj:1999gm}. They are based on asymmetric orbifolds and yield an {\em exactly} vanishing vacuum energy at 1-loop. However, contrary to what was initially believed, the 2-loop contribution seems to be non-trivial, as {\em a priori} expected \cite{L2}. It is important to stress that when these models describe a spontaneous breaking of supersymmetry to $\N=0$, they are super no-scale models in a strong sense and that, when perturbative heterotic dual descriptions are found, the latter appear to be super no-scale models in the weaker sense we have defined \ie {\em with boson-fermion classical degeneracy at the massless level only}  \cite{HarveyL=0,Angelantonj:1999gm}.

In Sect. \ref{snsm}, we display  one of the simplest super no-scale models. It is realized in heterotic string  compactified on $T^2\times T^2\times T^2$. The  moduli $T_2,U_2$ and $T_3,U_3$, associated to the $2^{\rm nd}$ and $3^{\rm rd}$ internal 2-tori, take values such that the right-moving gauge group is enhanced to either ${\cal G}=U(1)^2 \times SU(2)^4\times SO(16)^2$ or $U(1)^3\times SU(2)\times SU(3)\times SO(16)^2$. The  $\N=4\to 0$  spontaneous breaking of supersymmetry is realized \via  a  stringy Scherk-Schwarz mechanism \cite{SSstring} that involves the $1^{\rm st}$ 2-torus only, and the supersymmetry breaking scale $m_{3/2}$ is  a function of the associated moduli $T_1,U_1$. 

When $m_{3/2}$ is of the order of the string scale, a fact that arises when $\abs T_1\abs$ and $\abs U_1\abs$ are $\O(1)$, the corrections $\O(M_{\rm s}^4 \, e^{-c{M_{\rm s}/m_{3/2}}})$ to the effective potential are not suppressed anymore. Even if these precise terms are those responsible for Hagedorn-like transitions in models where supersymmetry is spontaneously broken to $\N=0$ \cite{Hage, Kounnas-Rostand}, we show that such instabilities are not present in our model. In other words, the theory does not develop classical tachyonic modes. Moreover, the super no-scale structure shows up  as soon as $m_{3/2}$ is lower than $M_{\rm s}$. This situation is encountered  in two distinct corners of the $(T_1,U_1)$-moduli space, which are T-dual to each other : $\abs T_1\abs\gg1$ with $\abs U_1\abs=\O(1)$, and $\abs T_1\abs=\O(1)$ with $\abs U_1\abs\ll1$.  On the contrary, $m_{3/2}$ is greater than $M_{\rm s}$ in the remaining corners of the $(T_1,U_1)$-moduli space, which are also T-dual to one another : $\abs T_1\abs\ll1$ with $\abs U_1\abs=\O(1)$, and $\abs T_1\abs=\O(1)$ with $\abs U_1\abs\gg1$. When $m_{3/2}>M_{\rm s}$, the model is naturally interpreted as an $\N=0$ theory realized as an {\em explicit} breaking of $\N=4$ (rather than a no-scale model).
It is also interesting to note that  when $m_{3/2}$ varies from  $+\infty$ to 0, $\V_{\mbox{\scriptsize 1-loop}}$ decreases monotonically and converges to~0. This behavior imposes the interesting fact that in a cosmological scenario, $m_{3/2}$ slides to  lower values, thus implying  the super no-scale structure to be reached dynamically at a low supersymmetry breaking scale. 

The above statement is valid provided that there are no  tachyonic instabilities, which can be developed at the 1-loop level. In order to study this issue, we consider in Sect.  \ref{deformations} the response of $\V_{\mbox{\scriptsize 1-loop}}$ under  all possible small moduli deformations of the $\Gamma_{6,6+16}$ lattice, namely the $T^6$-metric and antisymmetric tensor, and Wilson lines. The associated moduli $Y_{IJ}$, $\mbox{$I\in\{1,\dots, 6\}$}$, $J\in\{1,\dots,6+16\}$  cover  the full classical moduli space ${SO(6, 6+16)\over SO(6)\times SO(6+16)}$ around the initial extended symmetry point based on the gauge group  $U(1)^2\times SU(2)^4\times SO(16)^2$. Actually, slightly deforming the initial background amounts to switching on Higgs scales $Y_{IJ}M_{\rm s}$ smaller than $m_{3/2}$. In this case, some of the $n_{\rm B}+n_{\rm F}$ massless states acquire small masses. In fact, $n_{\rm B}$ and $n_{\rm F}$ are functions of the $Y_{IJ}$'s, which  actually interpolate between distinct integer values. Expanding locally around the initial background, we find 
\be
\label{vd}
\V_{\mbox{\scriptsize 1-loop}} =\xi(n_{\rm F}-n_{\rm B})\, m_{3/2}^4-\tilde\xi \, m_{3/2}^2 \sum_{\alpha}b_\alpha\!\! \sum_{J=1}^{{\rm rank}\, G_\alpha}\sum_{I=1}^6(Y_{IJ}M_{\rm s})^2+\cdots+\O\!\left(M_{\rm s}^4\, e^{-cM_{\rm s}/ m_{3/2}}\right) ,
\ee 
where $\tilde \xi>0$. The structure of this result happens to be valid for any no-scale model that realizes the $\N=4\to 0$ breaking of supersymmetry. The $G_\alpha$'s are the gauge group factors, and the $b_\alpha$'s are their associated $\beta$-function coefficients.  The $Y_{IJ}$'s are their Wilson lines along $T^6$. 
The above result  shows that the Wilson lines associated to  Cartan generators of an asymptotically free gauge group factor $G_\alpha$ ($b_\alpha <0$), acquire positive squared masses at $\mbox{1-loop}$ and  thus,  they are stabilized at the origin,  $Y_{IJ}=0$.  On the contrary, the moduli associated to a non-asymptotically free gauge group factor $G_\alpha$ ($b_\alpha >0$), become tachyonic. They condense, thus inducing negative contributions to  $\V_{\mbox{\scriptsize 1-loop}}$ and the Higgsing of $G_\alpha$ to subgroups with non-negative $\beta$-function coefficients but equal total rank.  It is only when $b_\alpha =0$ that the associated $Y_{IJ}$'s remain massless. 

Note however that the stability of  the super no-scale models is  always guaranteed  when they are considered at finite temperature $T$, as long as $T$ is greater than $m_{3/2}$. This follows from the fact that in the effective potential at finite temperature -- the quantum free energy~--, all squared masses are shifted by $T^2$, which implies that all moduli deformations are stabilized at $Y_{IJ}=0$ \cite{CosmologicalTerm}. Therefore, in a cosmological scenario where the Universe grows up and the temperature drops, the previously mentioned instabilities (for $b_\alpha >0$) take place as soon as $T^2$ reaches $m_{3/2}^2$ from above.   

In Sect. \ref{descendant},  we consider chains  of super no-scale models that realize an $\N=2\to 0$ or $\N=1\to 0$ spontaneous breaking of supersymmetry, \via $\Z^{\rm free}_2$ or $\Z^{\rm free}_2 \times \Z_2$ orbifold actions on parent $\N=4\to 0$ super no-scale models. In the ``descendant" theories, $\Z^{\rm free}_2$ is freely acting, which ensures that the  sub-breaking of $\N=4\to 2$ is {\it spontaneous}, so that the  models are free of decompactification problems \cite{N=0thresh}. The drawback of this chain of models is that the final spectrum is non-chiral, as opposed to that of the  super no-scale models based on non-freely acting orbifolds and constructed in Ref. \cite{ADM}, which however suffer from decompactification problems \cite{thresholds,universality,solving}.  

Finally, additional remarks and perspectives can be found in Sect. \ref {cl}. 


\section{\bm $\N=4\to 0$ super no-scale model}
\label{snsm}

In this section, we built and analyze in more details one of the simplest super no-scale models, already presented in Ref. \cite{planck2015}. It is constructed in heterotic string and realizes the $\N=4\to 0$ spontaneous  supersymmetry breaking, with gauge symmetry that  will appear to be either  ${\cal G}=U(1)^2\times SU(2)^4\times SO(16)^2$ or $U(1)^3\times SU(2)\times SU(3)\times SO(16)^2$.  The  1-loop effective potential is given as usual in terms of the   partition function at genus 1, $Z_{\rm sss}$, integrated over the fundamental domain $\F$ of $SL(2,\Z)$,
\be
\label{vi}
\V_{\mbox{\scriptsize 1-loop}}= -{M_{\rm s}^4\over (2\pi)^4}\int_\F {d^2\tau\over 2\tau_2^2}\, Z_{\rm sss}\, ,
\ee 
where $\tau= \tau_1+i\tau_2$ is the genus-1 Techm\"uller parameter.


\subsection{Partition function}
\label{spectrumG}
Our starting point is the ``parent" $\N=4$, $~E_8\times E_8'$ heterotic string compactified on $T^2\times T^2\times T^2$, whose partition function has the following factorized  form :
\be
\label{N4}
Z_{\N=4}= O^{(0)}_{2,2}\;   O^{(1)}_{2,2}\;  O^{(2)}_{2,2}\;  O^{(3)}_{2,2}\; {1\over 2}\sum_{a,b}Z^{(\rm F)}_{4,0}\!\big[{}^{a}_{b}  \big] \,
 Z_{0, 8+8}\, ,
\ee
 where 
 $Z_{4,0}^{(\rm F)}\!\big[{}^{a}_{b}  \big]$ denotes  the contribution of the  left-moving  2-dimensional fermions, super-partners of the $2+6$ coordinates in light-cone gauge,  and  $Z_{0,8+8}$ is that   of   the  $8 + 8$  right-moving compact bosons,  which give rise to the $E_8\times E'_8$ affine characters in the adjoint representation,
\be
Z_{4,0}^{(\rm F)}\!\big[{}^{a}_{b}  \big]\!=(-1)^{a+b+ab}\, {\theta\big[{}^a_b\big]^4\over \eta^4}\, , \qquad Z_{0,8+8}=\!\left({1\over 2}\sum_{\gamma,\delta} {\bar \theta\big[{}^\gamma_\delta\big]^8\over \bar \eta^8}\,\right) \!\!\left( {1\over 2}\sum_{\gamma',\delta'} {\bar \theta\big[{}^{\gamma'}_{\delta'}\big]^8\over \bar \eta^8}\,\right) ,
\ee
where the spin structure $a,b$ and  $\gamma,\delta,\gamma',\delta'\in \Z_2$.

$O_{d-2,d-2}^{(0), }$ denotes the contributions of the $d-2=2$ spacetime light-cone coordinates, while  
$O_{2,2}^{(I)}$, $I\in\{1,2,3\}$,  arise from the coordinates of the three internal 2-tori and can be expressed in terms  lattices :
\be
O^{(0)}_{d-2,d-2} ={1\over \left(\sqrt{\tau_2}\eta \bar \eta\right)^{d-2}}\, , \qquad \quad O^{(I)}_{2,2}={\Gamma_{2,2(T_I,U_I)}\over\eta^2\bar \eta^2} \, , \;\;I\in\{1,2,3\}\,.
\ee
We denote by  $\Gamma_{2,2}=\Gamma_{2,2}\big[{}^{0,\, 0}_{0,\, 0} \big]$ the unshifted  $(2,2)$-lattice. More generally, the shifted lattice to be used in a moment is defined as $\Gamma_{2,2}\big[{}^{h_1,\, h_2}_{g_1\, ,\, g_2} \big]$, where we limit ourselves to shifts $h_1,g_1$ and $h_2,g_2\in\Z_2$, 
\be
 \label{lat}
\Gamma_{2,2}\big[{}^{h_1,\, h_2}_{g_1\, ,\, g_2} \big]\!_{(T,U)} =\sum_{\scriptsize\begin{array}{c}m_1,m_2\\ n_1,n_2\end{array}}\!\!\!e^{i\pi ({g_1 m_1+g_2 m_2})}\, q^{{1\over 2}\abs p_L\abs^2}\bar q^{{1\over 2}\abs p_R\abs^2}\, ,
\ee
where $q=e^{2i\pi\tau}$ and
\begin{align}
p_L&={1\over \sqrt{2\, \Im T\, \Im U}}\left[Um_1-m_2+T \big(n_1+{1\over 2}h_1\big)+T U \big(n_2+{1\over 2}h_2\big)\right],\nonumber\\
p_R&={1\over \sqrt{2\, \Im T\, \Im U }}\left[ Um_1-m_2+\bar T\big(n_1+{1\over 2}h_1\big)+\bar T U \big(n_2+{1\over 2}h_2\big)\right].
\end{align}
$T_I$ and $U_I$ are given as usual in terms of the internal metric $G_{ij}$ and antisymmetric tensor $B_{ij}$, $i,j\in\{1,\dots,6\}$,
\begin{align}
&T_I=i\sqrt{G_{2I-1,2I-1}G_{2I,2I}-G_{2I-1,2I}^2}+B_{2I,2I-1} \, , \nonumber \tesp\\
& U_I={i\sqrt{G_{2I-1,2I-1}G_{2I,2I}-G_{2I-1,2I}^2}+G_{2I,2I-1}\over G_{2I-1,2I-1}}\, , \quad I\in\{1,2,3\}\, .
\end{align}
In the above expressions, $\theta\big[{}^a_b\big]\!(\nu\abs\tau)$ (or $\theta_{\alpha}(\nu\abs\tau)$, $\alpha\in\{1,2,3,4\}$, to be used later) are the Jacobi elliptic functions and $\eta$ is the Dedekind function, following the conventions of Ref.  \cite{KiritsisBook}.

It is also convenient to introduce the $O(2N)$ characters defined as
\begin{align}
O_{2N}&={\theta\big[{}^0_0\big]^N+\theta\big[{}^0_1\big]^N\over 2\eta^N}\, , &V_{2N}&={\theta\big[{}^0_0\big]^N-\theta\big[{}^0_1\big]^N\over 2\eta^N}\, ,\nonumber \tesp\\
S_{2N}&={\theta\big[{}^1_0\big]^N+(-i)^N\theta\big[{}^1_1\big]^N\over 2\eta^N}\, ,
&C_{2N}&={\theta\big[{}^1_0\big]^N-(-i)^N\theta\big[{}^1_1\big]^N\over 2\eta^N}\, ,
\label{charac}
\end{align}
in terms of which  we can write  $Z_{\N=4}$ in the following  factorized form,
\be
Z_{\N=4}= O^{(0)}_{2,2}\;  O^{(1)}_{2,2}\; O^{(2)}_{2,2}\; O^{(3)}_{2,2} \;\big( V_8-S_8\big)\!\left(\bar O_{16}+\bar S_{16}\right) \!\left( \bar O'_{16}+\bar S'_{16}\right),
\ee
where the $E_8$ character becomes $\bar O_{16}+\bar S_{16}$. 

We then  introduce a stringy Scherk-Schwarz mechanism \cite{SSstring} that  simultaneously  breaks $\N=4\to 0$ and $E_8\times E'_8\to SO(16)\times SO(16)'$, spontaneously. This is done by implementing a $\Z_2^{\rm shift}$ orbifold action that shifts the $1^{\rm st}$ internal direction, $X^1$. The associated lattice shifts $h,g\in\Z_2$ are coupled to the spin structure \via a non-trivial sign $S_L$, as well as to the $SO(16)$ and $SO(16)'$ spinorial characters with another sign $S_R$. In total, this amounts to replacing  
\begin{align}
\label{sign}
O_{2,2}^{(1)}&\longrightarrow {1\over 2}\sum_{h,g}S_L\big[{}^{a;\, h} _{b;\;  g} \big]\,{\Gamma_{2,2}\big[{}^{h,\, 0}_{g,\, 0} \big]\!_{(T_1,U_1)}\over \eta^2\bar\eta^2}&\with &&S_L\big[{}^{a;\, h} _{b;\;  g} \big]&=(-1)^{ga+hb+hg}\, ,\nonumber \\
Z_{0,16}&\longrightarrow  {1\over 2}\sum_{\gamma,\delta}\, {1\over 2}\sum_{\gamma',\delta'} S_R\big[{}^{\gamma,\, \gamma';\, h}_{\delta,\, \delta'\, ;\, g} \big]\,{\bar \theta\big[{}^\gamma_\delta\big]^8\over \bar \eta^8}\, {\bar \theta\big[{}^{\gamma'}_{\delta'}\big]^8\over \bar \eta^8}&\with && S_R\big[{}^{\gamma,\, \gamma';\, h}_{\delta,\, \delta'\, ;\; g} \big]&=(-1)^{g(\gamma+\gamma')+h(\delta+\delta')}\, .
\end{align}
The shift $g$ being coupled by the sign $S_LS_R$ to the spacetime fermions ($a=1$), to the $SO(16)$ spinorial characters ($\gamma=1$) and to the  $SO(16)^{\prime}$ spinorial characters ($\gamma'=1$), the model will be referred as ``spinorial-spinorial-spinorial", or sss-model. Its partition function  is
\begin{align}
\label{sss1}
Z_{\rm sss}= &\; O^{(0)}_{2,2}\; O^{(2)}_{2,2}\; O^{(3)}_{2,2}\;  {1\over 2}\sum_{h,g}{\Gamma_{2,2}\big[{}^{h,\, 0}_{g,\, 0} \big]\!_{(T_1,U_1)}\over \eta^2\bar\eta^2}\;\times
\nonumber\\
 &\; {1\over 2}\sum_{a,b}(-1)^{a+b+ab}\, {\theta\big[{}^a_b\big]^4\over \eta^4}\, (-1)^{ga+hb+hg}\; \;{1\over 2}\sum_{\gamma,\delta}{\bar \theta\big[{}^\gamma_\delta\big]^8\over \bar \eta^8}\, (-1)^{g\gamma+h\delta}
\; \;{1\over 2}\sum_{\gamma',\delta'}{\bar \theta\big[{}^{\gamma'}_{\delta'}\big]^8\over \bar \eta^8}\, (-1)^{g\gamma'+h\delta'}\, ,
\end{align}
which leads  to
\begin{align}
Z_{\rm sss}=O^{(0)}_{2,2} \, O^{(2)}_{2,2}\,  O^{(3)}_{2,2} \,   {1\over2\eta^2\bar\eta^2} \, &\!\Big[\, \;\Gamma_{2,2}\big[{}^{0,\, 0}_{0,\, 0} \big]\!_{(T_1,U_1)}\left( V_8-S_8\right)\left(\bar O_{16}+\bar S_{16}\right) \left( \bar O'_{16}+\bar S'_{16}\right)\nonumber \\
 &\!\!\!\!\;+\Gamma_{2,2}\big[{}^{0,\, 0}_{1,\, 0} \big]\!_{(T_1,U_1)}\left(V_8+S_8\right)\left(\bar O_{16}-\bar S_{16}\right) \left(\bar O'_{16}-\bar S'_{16}\right)\nonumber \\
&\!\!\!\!\;+\Gamma_{2,2}\big[{}^{1,\, 0}_{0,\, 0} \big]\!_{(T_1,U_1)}\left(O_8-C_8\right)\left(\bar V_{16}+\bar C_{16}\right) \left( \bar V'_{16}+\bar C'_{16}\right)\nonumber \\
&\!\!\!\!\;-\Gamma_{2,2}\big[{}^{1,\, 0}_{1,\, 0} \big]\!_{(T_1,U_1)}\left(O_8+C_8\right)\left(\bar V_{16}-\bar C_{16}\right) \left( \bar V'_{16}-\bar C'_{16}\right)\, \Big].
\end{align}
Defining the characters of the shifted $(2,2)$-lattice associated to the $1^{\rm st}$ 2-torus as 
 \be
 \label{defO}
O^{(1)}_{2,2}\big[{}^h_g\big]={\Gamma_{2,2}\big[{}^{h,\, 0}_{0,\, 0} \big]\!_{(T_1,U_1)}+(-1)^g\, \Gamma_{2,2}\big[{}^{h,\, 0}_{1,\, 0} \big]\!_{(T_1,U_1)}\over 2\eta^2\bar\eta^2}\, ,
 \ee
 the partition function of the sss-model takes the final form
 \begin{align}
\label{Zsss}
Z_{\rm sss}=O^{(0)}_{2,2} \; O^{(2)}_{2,2}\;  O^{(3)}_{2,2} \, &\Big[\; \,O_{2,2}^{(1)}\big[{}^0_0\big]\Big( V_8(\bar O_{16}  \bar O'_{16}+ \bar S_{16} \bar S'_{16} ) -S_8 (\bar O_{16}\bar S'_{16} + \bar S_{16} \bar O'_{16})\Big) \nonumber \\
& \!\!+O_{2,2}^{(1)}\big[{}^0_1\big]\Big( V_8 (\bar O_{16}\bar S'_{16} + \bar S_{16} \bar O'_{16}) -S_8  (\bar O_{16} \bar O'_{16}+ \bar S_{16}\bar S'_{16} ) \Big)\nonumber\\
&\!\!+O_{2,2}^{(1)}\big[{}^1_0\big]\Big( O_{8} (\bar V_{16} \bar C'_{16}+ \bar C_{16}\bar V'_{16} ) -C_8 (\bar V_{16}\bar V'_{16} + \bar C_{16}\bar C'_{16} )\Big) \nonumber\\
& \!\!+O_{2,2}^{(1)}\big[{}^1_1\big]\Big( O_8(\bar V_{16}\bar V'_{16} + \bar C_{16}\bar C'_{16}) -C_8 (\bar V_{16} \bar C'_{16}+ \bar C_{16}\bar  V'_{16} )\Big)\, \Big].
 \end{align}
 
 For comparison, we also display  the model where only $S_L$ is introduced ($S_R\equiv 1$). The latter realizes the $\N=4 \to 0$ breaking but preserves the full $E_8\times E_8'$ gauge symmetry. Since in that case the shift $g$ is only coupled to the spacetime fermions ($a=1$), this model will be referred as ``spinorial'',  or s-model. The associated partition function is
\be
Z_{\rm s}= O^{(0)}_{2,2} \; O^{(2)}_{2,2} \; O^{(3)}_{2,2}
\left(O^{(1)}_{2,2}\big[{}^0_0\big]V_8 - O^{(1)}_{22}\big[{}^0_1\big]S_8 - O^{(1)}_{22}\big[{}^1_0\big]C_8 + O^{(1)}_{22}\big[{}^1_1\big]O_8\right) \!
\left(\bar O_{16}+\bar S_{16}\right) \!\left(\bar O'_{16}+\bar S'_{16}\right)\!,
\ee
with factorized right-moving characters.  $Z_{\rm s}$ is similar to the partition function of the initial $\N=4$ model at finite temperature \cite{Kounnas-Rostand,CosmologicalTerm}. The latter is obtained  by replacing the role of the $1^{\rm st}$ internal direction $X^1$ with that of a compact Euclidean time $X^0$ of perimeter $\beta=2\pi R_0=M_{\rm s}/T$, where $T$ is the temperature. 

The spectra of the s- and sss-model can be easily studied by observing that  the $1^{\rm st}$ 2-torus characters can be written as
\be
O^{(1)}_{2,2}\big[{}^h_g\big]={1\over \eta^2\bar \eta^2}\!\!\!\sum_{\scriptsize\begin{array}{c}k_1,m_2\\ n_1,n_2\end{array}}\!\!\!q^{{1\over 2}\abs p^{(1)}_L\abs^2}\, \bar q^{{1\over 2}\abs p^{(1)}_R\abs^2} ,
\ee
where the momentum $m_1$ is redefined as $2k_1+g$,
\begin{align}
\label{pLR}
p_L^{(1)}&={1\over \sqrt{2\, \Im T_1\, \Im U_1}}\left[U_1(2k_1+g)-m_2+{T_1\over 2}\big(2n_1+h\big)+T_1U_1 n_2\right],\nonumber\\
p_R^{(1)}&={1\over \sqrt{2\, \Im T_1\, \Im U_1}}\left[U_1(2k_1+g)-m_2+{\bar T_1\over 2}\big(2n_1+h\big)+\bar T_1 U_1 n_2\right].
\end{align}
$~$\\
In particular, the scale  $m_{3/2}$ of  $\N=4\to 0$ spontaneous  supersymmetry breaking satisfies  
\begin{equation}
\label{m32}
m^2_{3/2}={\abs U_1\abs^2 M^2_{\rm s}\over \Im T_1\, \Im U_1}\, .
\end{equation} 
In the s-model, the sector $O^{(0)}_{2,2} O^{(1)}_{2,2}\big[{}^1_1\big] O^{(2)}_{2,2} O^{(3)}_{2,2}   O_8 \bar O_{16} \bar O'_{16}$ contains tachyonic states when the supersymmetry breaking scale $m_{3/2}$ is of order $M_{\rm s}$. In this case, the integrated partition function \ie the effective potential is ill-defined and a Hagedorn-like instability actually arises \cite{Hage, Kounnas-Rostand}. In the $\N=4$ theory at finite temperature, this phenomenon is nothing but the well known Hagedorn instability,  which takes place when $\sqrt{2}\, (\sqrt{2}-1)<R_0<\sqrt{2}\, (\sqrt{2}+1)$.
On the contrary, the situation happens to be drastically different in the sss-model. The reason is that the sector with reversed GSO projection, which is characterized by the left-moving character $O_8$, is dressed by right-moving characters that start at the massless level, $\bar V_{16}\bar V'_{16}$. Therefore, the level matching condition prevents any physical tachyon to arise for arbitrary $T_I,U_I$, $\{I=1,2,3\}$. No Hagedorn-like instability occurs and the 1-loop effective potential based on the partition function $Z_{\rm sss}$ is well defined. 

However, marginal deformations other than $T_I,U_I$ can be switched on. Beside the dilaton, the classical moduli space can be parameterized  by the 6 scalars of the bosonic degrees of freedom of the $\N=4$ vector multiplets that realize the $U(1)^{6+16}$ Cartan gauge symmetry (the fermionic superpartners are massive). It takes the form   
\be
{SU(6)\times SO(6+16)\over SO(6)\times SO(16)}
\label{modspace}
\ee
and its dimension is $6\times (6+16)$. For small enough deformations away from the sss-model, tachyonic instabilities would not arise. On the contrary, some $\O(1)$ Wilson lines deformations can certainly lead to tachyonic modes, when the gravitino mass is of order $M_{\rm s}$ \cite{GV}. Note however that theories where all potentially dangerous moduli deformations have been projected out  do exist, as shown explicitly in a four-dimensional orientifold model constructed in Ref. \cite{Angelantonj:2006ut}. 

Before concluding this subsection, we give the expression of the 1-loop effective potential of the s- and sss-model, when $\Im T_1\gg 1$ and $U_1=\O(i)$, which implies $m_{3/2}\ll M_{\rm s}$ \cite{N=0thresh}. As we will be seen in details in Sect. \ref{deformations}, $\V_{\mbox{\scriptsize 1-loop}}$ takes in this regime the following form :
\be
\label{v1loop}
\V_{\mbox{\scriptsize 1-loop}}= {n_{\rm F}-n_{\rm B}\over 16\pi^7}\, {M_{\rm s}^4\over (\Im T_1)^2}\, E_{(1,0)}(U_1\abs 3,0)+{\cal O}\!\left(M_{\rm s}^4\, e^{-c\sqrt{\Im T_1}}\right),
\ee
where $n_{\rm F}$ and $n_{\rm B}$ are the numbers of fermionic and bosonic massless degrees of freedom\footnote{The factor $c>0$ appearing in the exponentially suppressed terms depends on all moduli but $\Im T_1$ and the dilaton. It is of order $M/M_{\rm s}$, where $M$ is the lowest mass above the pure KK mass scale  $m_{3/2}$. In the s- and sss-model, it is of order $M_{\rm s}$, but can be in other cases a large Higgs scale or GUT scale (See Sect. \ref{deformations}).},  and the functions 
\be
E_{(g_1,g_2)}(U\abs  s,k)={\sum_{\tilde m_1,\tilde m_2}}^{\!\!\!\prime}{(\Im U)^s\over \left(\tilde m_1+{g_1\over 2}+(\tilde m_2+{g_2\over 2})U\right)^{s+k}(\tilde m_1+{g_1\over 2}+\left(\tilde m_2+{g_2\over 2})\bar U\right)^{s-k}}
\ee
are shifted complex Eisenstein series of asymmetric weights, where $g_1,g_2\in \Z_2$.  
While $n_{\rm F}=0$ for the s-model  and 
$\V_{\mbox{\scriptsize 1-loop}}$ scales like $m_{3/2}^4$, we are going to see that the sss-model can be super no-scale.  


\subsection{\bm The super no-scale regime, $m_{3/2}\ll M_{\rm s}$}
\label{snsRegime}
In order to show that the 1-loop effective potential of the sss-model can be exponentially suppressed, $\O(M_{\rm s}^4\, e^{-c\sqrt{\Im T_1}})$, when the supersymmetry breaking scale is low, we look for conditions such that the massless fermions and bosons present in the regime $\Im T_1 \gg1$, $U_1=\O(i)$ satisfy $n_{\rm F}= n_{\rm B}$ \cite{planck2015}. 

Given the fact that the states in the sectors $O^{(1)}_{2,2}\big[{}^1_g\big]$, $g=0,1$, have non-trivial winding numbers $2k_1+1$ along the very large compact direction $X^1$, they are super massive. In order to find the massless (or more generally light) states of the sss-model, it is only required to analyze the sectors $O^{(1)}_{2,2}\big[{}^0_g\big]$, $g=0,1$. 

\vspace {0,3cm}
\noindent {\large \em Sector $O^{(1)}_{2,2}\big[{}^0_0\big]\!_{(T_1,U_1)}$}

\noindent The bosonic sector $O^{(0)}_{2,2}  O^{(1)}_{2,2}\big[{}^0_0\big]  O^{(2)}_{2,2} O^{(3)}_{2,2}   V_8 \bar O_{16} \bar O'_{16}$ contains massless degrees of freedom, which are associated to the graviton, antisymmetric tensor, moduli fields (dilaton, Wilson lines, internal metric and antisymmetric tensor) and to a vector boson  in the adjoint representation of a gauge group ${\cal{G}}=G^{(1)}\times G^{(2)}\times G^{(3)}\times SO(16)\times SO(16)'$, where the factor $G^{(I)}$ arises from the lattice associated to the $I^{\rm th}$ $\mbox{2-torus}$. In the regime we consider, $G^{(1)}=U(1)^2$ but $G^{(I)}$, $I\in\{2,3\}$, may be a higher dimensional group of rank 2. For generic $T_I,U_I$, $I\in\{2,3\}$, we have   $G^{(I)}=U(1)^2$, which can be enhanced to $SU(2)\times U(1)$, $SU(2)^2$ or $SU(3)$ at particular points in moduli space. The degeneracy of these massless states is
\begin{align}
n_{\rm B}\equiv d(\mbox{Bosons}\big[{}^0_0\big])&= d(V_8)\!\left[d(O^{(0)}_{2,2})+d(O^{(1)}_{2,2}\big[{}^0_0\big])+d(O^{(2)}_{2,2}) + d(O^{(3)}_{2,2})+d(\bar O_{16})+d(\bar O'_{16})\right]\nonumber \\
&=8\times \big[2+2+d(G^{(2)}) + d(G^{(3)}) +8\times 15+ 8\times 15  \big] \nonumber \\
&=\underline{8\times \big[244+d(G^{(2)})+d(G^{(3)})\big]} ,
\end{align}
which depends on the moduli $T_I,U_I$, $I\in\{2,3\}$.

Similarly, the fermionic sector $-O^{(0)}_{2,2}  O^{(1)}_{2,2}\big[{}^0_0\big]  O^{(2)}_{2,2} O^{(3)}_{2,2}   S_8 (\bar O_{16} \bar S'_{16}+\bar S_{16} \bar O'_{16})$ begins at the massless level, with states in the spinorial representations of $SO(16)$ or $SO(16)^{\prime}$. Their multiplicity is
\begin{equation}
n_{\rm F}\equiv d(\mbox{Fermions}\big[{}^0_0\big])= d(S_8)\!\left[d(\bar S'_{16})+d(\bar S_{16})\right]=8\times (128+128)=\underline{8\times 256}\, ,
\end{equation}
which is independent of the point  in moduli space we sit at. Moreover, the above bosonic and fermionic degrees of freedom are accompanied by light towers of pure KK states associated to  the  $1^{\rm st}$ 2-torus. Their momenta along the directions $X^1$ and $X^2$, which are both large, are $2k_1$ and $m_2$, and their KK masses are of order $m_{3/2}$. 

\vspace {0,3cm}
\noindent {\large \em  Sector $O^{(1)}_{2,2}\big[{}^0_1\big]\!_{(T_1,U_1)}$}

\noindent The bosonic sector $O^{(0)}_{2,2}  O^{(1)}_{2,2}\big[{}^0_1\big]  O^{(2)}_{2,2} O^{(3)}_{2,2} V_8 (\bar O_{16} \bar S'_{16}+\bar S_{16} \bar O'_{16})$ contains light towers of KK modes arising from the $1^{\rm st}$ 2-torus. Their momenta along $X^1$ and $X^2$ are $2k_1+1$ and $m_2$, the oddness of the former implying they cannot be massless. Their degeneracy is 
 \begin{equation}
d({\rm Bosons}\big[{}^0_1\big])= d(V_8)\!\left[d(\bar S'_{16})+d(\bar S_{16})\right] =\underline{8\times 256}\, ,
\end{equation} 
which equals $n_{\rm F}$.

Similarly, the fermionic sector $-O^{(0)}_{2,2}  O^{(1)}_{2,2}\big[{}^0_1\big]  O^{(2)}_{2,2} O^{(3)}_{2,2} S_8 \bar O_{16} \bar O'_{16}$ contains light KK states, with non-vanishing masses, their momenta being again $2k+1$ and $m_2$.  Their counting goes as follows :  
\begin{align}
d({\rm Fermions}\big[{}^0_1\big])&= d(S_8)\!\left[d(O^{(0)}_{2,2})+d(O^{(1)}_{2,2}\big[{}^0_1\big])+d(O^{(2)}_{2,2}) + d(O^{(3)}_{2,2})+d(\bar O_{16})+d(\bar O'_{16})\right]\nonumber \\
&=8\times \big[2+2+d(G^{(2)}) + d(G^{(3)}) +8\times 15+ 8\times 15  \big] \nonumber \\
&=\underline{8\times \big[244+d(G^{(2)})+d(G^{(3)})\big]} ,
\end{align}
which equals $n_{\rm B}$. 

The fact that the number of KK towers with odd momenta equals that of those with even momenta is not a coincidence. In the initial $\N=4$ theory, among the characters with even $\gamma+\gamma'$, those corresponding to spacetime fermions are given a KK mass in the sss-model, while those associated to spacetime bosons are not modified. This feature is common to the s-model,
\be
O_{2,2}^{(1)}(V_8-S_8)(\bar O_{16}  \bar O'_{16}+ \bar S_{16} \bar S'_{16} )\longrightarrow \left(O_{2,2}^{(1)}\big[{}^0_0\big] V_8-O_{2,2}^{(1)}\big[{}^0_1\big] S_8 \right)\!(\bar O_{16}  \bar O'_{16}+ \bar S_{16} \bar S'_{16} )\, .
\ee
On the contrary, when $\gamma+\gamma'$ is odd, the sign $S_R$ effectively reverses the roles of bosons and fermions. Among the characters with odd $\gamma+\gamma'$, those corresponding to spacetime bosons are given a KK mass in the sss-model, while those associated to spacetime fermions are not modified. These facts are opposite to those encountered in the the s-model. The sss case thus leads 
\be
O_{2,2}^{(1)}(V_8-S_8)(\bar O_{16}  \bar S'_{16}+ \bar S_{16} \bar O'_{16} )\longrightarrow \left(O_{2,2}^{(1)}\big[{}^0_1\big] V_8-O_{2,2}^{(1)}\big[{}^0_0\big] S_8 \right)\!(\bar O_{16}  \bar S'_{16}+ \bar S_{16} \bar O'_{16} )\, .
\ee

The condition for the sss-model to be super no-scale is that the numbers of massless fermions and bosons be equal,
\begin{equation}
n_{\rm F}=n_{\rm B}\, ;\qquad  n_{\rm F}=8\times 256\; ,  \qquad n_{\rm B}=8\times \big[244+d(G^{(2)})+d(G^{(3)})\big].
\end{equation} 
 This imposes \cite{planck2015} $\, d(G^{(2)})+d(G^{(3)})=12$, which leads for $\rk(G^{(2)})=\rk(G^{(3)})=2$,
\be
(a) \;\;G^{(2)}\times G^{(3)}=SU(2)^4\qquad \mbox{or} \qquad (b) \;\;G^{(2)}\times G^{(3)}=SU(3)\times SU(2)\times U(1)\, .
\ee
Modulo T-duality, Solution $(a)$ is realized at the self-dual point  $T_2=U_2=T_3=U_3=i$, which leads the enhanced $G^{(2)}\times G^{(3)}=SU(2)^4=SO(4)^2$ gauge symmetry. Note that in the neighborhood of this point, some of the $SU(2)$ factors are spontaneously broken to $U(1)$. In this case, $n_{\rm B}$ takes lower values and $\V_{\mbox{\scriptsize 1-loop}}$, given in Eq. (\ref{v1loop}), becomes positive. Thus, at the above self-dual point, the 1-loop effective potential is positive semi-definite with respect to the variables $T_I,U_I$, $I\in\{1,2,3\}$, where $\Re T_1$, $m_{3/2}$ and $U_1$ are flat directions. The moduli $T_I,U_I$, $I\in\{2,3\}$, are attracted dynamically to the self-dual point, which is characterized by a super no-scale structure. In Sect. \ref{deformations}, we will consider in great details all moduli deformations, locally around Background ($a$), and the associated response of the effective potential. 

Solution $(b)$ occurs modulo T-duality at $T_2=U_2=e^{i\pi/3}$, $T_3=U_3$ arbitrary. Locally around this complex line, $G^{(2)}\times G^{(3)}$ is spontaneously broken to a subgroup and $n_{\rm B}$ decreases. Thus, the 1-loop effective potential is locally positive semi-definite with respect to $T_I,U_I$, $I\in\{1,2,3\}$, where the flat directions are parameterized by $\Re T_1$, $m_{3/2}$, $U_1$ and $T_3=U_3$.   
Again, the model is naturally super no-scale; the trajectories of the  time-dependent moduli associated to the $2^{\rm nd}$ and $3^{\rm rd}$ 2-tori being attracted to these points.  


\subsection{The T-dual regimes}
\label{DualRegime}
We have seen that for $T_1\to i\infty$, $U_1=\O(i)$, the sss-model is characterized by a low supersymmetry breaking scale $m_{3/2}$ and a super no-scale structure. In the present subsection, our goal is to study the remaining corners of the moduli space where  either $T_1$ or $U_1$ (but not both) is of order $i$. We thus define 4 regimes, 
\begin{align}
(\mbox{I})&:\;  T_1\to i\infty\, , \; U_1=\O(i)\nonumber\\
(\mbox{II})&:\;  T_1\to 0\, , \; U_1=\O(i)\nonumber\\
(\mbox{III})&:\;  T_1=\O(i)\, , \; U_1\to i\infty\nonumber\\
(\mbox{IV})&:\;  T_1=\O(i)\, , \; U_1\to 0\, ,
\end{align}
where the first one is super no-scale with $m_{3/2}<M_{\rm s}$, while the others can be respectively analyzed by defining T-dual moduli,
\begin{align}
\hspace{1cm}(\mbox{II})&:\;  (\hat T_1,\hat U_1)=\left(-{2\over T_1},-{1\over 2U_1}\right)\nonumber\\
(\mbox{III})&:\;  (\check T_1,\check U_1)=\left(2U_1,{T_1\over 2}\right)\nonumber\\
(\mbox{IV})&:\;  (\tilde T_1,\tilde U_1)=\left(-{1\over U_1},-{1\over T_1}\right) .
\end{align}
In terms of these new variables, Regime (II) is reached by taking $\hat T_1\to i\infty$, $\hat  U_1=\O(i)$, Regime (III) corresponds to $\check T_1\to i\infty$, $\check U_1=\O(i)$, and Regime (IV) is associated to $\tilde T_1\to i\infty$, $\tilde  U_1=\O(i)$. The relevance of the above definitions of T-dual moduli follows from the fact that  
\be
\label{Tdual}
O_{2,2}^{(1)}\big[{}^h_g\big]\!_{(T_1,U_1)}=O_{2,2}^{(1)}\big[{}^g_h\big]\!_{(\hat T_1,\hat U_1)}=O_{2,2}^{(1)}\big[{}^g_h\big]\!_{(\check T_1,\check U_1)}=O_{2,2}^{(1)}\big[{}^h_g\big]\!_{(\tilde T_1,\tilde U_1)}\, .
\ee
The third equality is telling us that the sss-model (as well as the s-model) is self-dual  under the T-duality transformation $(T_1,U_1)\to (\tilde T_1,\tilde U_1)$,
\be
\label{Zdual}
Z_{\rm sss}(T_1,U_1)=Z_{\rm sss}(\tilde T_1,\tilde U_1)\, . 
\ee
Thus, the corners (I) and (IV) of the $1^{\rm st}$ 2-torus moduli space share a common behavior~: The sss-model is super no-scale in both limits, and the supersymmetry breaking scale satisfies
\be
m^2_{3/2}={\abs U_1\abs^2 M^2_{\rm s}\over \Im T_1\, \Im U_1}\ll M_{\rm s}^2\quad \mbox{in Regimes (I) and (IV)}\, ,
\ee 
which is a T-duality invariant expression.
On the contrary, the $1^{\rm st}$ equality in Eq. (\ref{Tdual}) allows us to rewrite the partition function as
 \begin{align}
\label{ZsssDual}
Z_{\rm sss}(T_1,U_1)&= \hat Z_{\rm sss}(\hat T_1,\hat U_1)&\nonumber \\
&=O^{(0)}_{2,2} \; O^{(2)}_{2,2} \;O^{(3)}_{2,2}\!\!\!\!&&\Big[\; \,O_{2,2}^{(1)}\big[{}^0_0\big]\!_{(\hat T_1,\hat U_1)}\Big( V_8(\bar O_{16}  \bar O'_{16}+ \bar S_{16} \bar S'_{16} ) -S_8 (\bar O_{16}\bar S'_{16} + \bar S_{16} \bar O'_{16})\Big) \nonumber \\
&&& \!\!+O_{2,2}^{(1)}\big[{}^0_1\big]\!_{(\hat T_1,\hat U_1)}\Big( O_{8} (\bar V_{16} \bar C'_{16}+ \bar C_{16}\bar V'_{16} ) -C_8 (\bar V_{16}\bar V'_{16} + \bar C_{16}\bar C'_{16} )\Big)\nonumber\\
&&&\!\!+O_{2,2}^{(1)}\big[{}^1_0\big]\!_{(\hat T_1,\hat U_1)}\Big( V_8 (\bar O_{16}\bar S'_{16} + \bar S_{16} \bar O'_{16}) -S_8  (\bar O_{16} \bar O'_{16}+ \bar S_{16}\bar S'_{16} )\Big)\nonumber\\
&&&\!\!+O_{2,2}^{(1)}\big[{}^1_1\big]\!_{(\hat T_1,\hat U_1)}\Big( O_8(\bar V_{16}\bar V'_{16} + \bar C_{16}\bar C'_{16}) -C_8 (\bar V_{16} \bar C'_{16}+ \bar C_{16}\bar  V'_{16} )\Big)\, \Big],
\end{align}
which shows that the sss-model is not self-dual under the T-duality transformation $(T_1,U_1)\to (\hat T_1,\hat U_1)$. Note that in the s-model, this transformation amounts to inter-exchanging the spinorial characters $S_8\leftrightarrow C_8$ \ie reversing spacetime chirality. The latter being a matter of convention, Regimes (II) and (III) describe isomorphic particle contents in the s-model.  

Finally, the $2^{\rm nd}$ equality in Eq. (\ref{Tdual}) guaranties the sss-model (as well as the s-model) is T-duality invariant under the transformation $(\hat T_1,\hat U_1)\to (\check T_1, \check U_1)$, which is nothing but the already mentioned symmetry  
$(T_1,U_1)\to (\tilde T_1,\tilde U_1)$. In other words, the identity (\ref{Zdual}) can be rewritten as 
\be
\hat Z_{\rm sss}(\hat T_1,\hat U_1)=\hat Z_{\rm sss}(\check T_1,\check U_1)\, . 
\ee
The above expression guaranties that the corners (II) and (III)  of the $1^{\rm st}$ 2-torus moduli space yield a common behavior. In the following, we describe the light spectrum and effective potential in these regimes. 

The winding numbers along the directions of the T-dual 2-torus whose \Ka and complex structure are $\hat T_1$ and $\hat U_1$ are $2k_1+g$ and $m_2$, which implies that in Regime (II), where $\Im \hat T_1\gg 1$, $\hat U_1=\O(i)$, the states with non-vanishing $2k_1+1$ or $m_2$ are super massive. Therefore, the pure T-dual KK modes lead exponentially dominant contributions, as follows from the expression of the T-dual 2-torus characters in Regime (II), which for $g=0$ are
\begin{align}
\label{O1}
O_{2,2}^{(1)}\big[{}^0_h\big]\!_{(\hat T_1,\hat U_1)}&={1\over \eta^2\bar\eta^2}\sum_{n_1,n_2}(q\bar q)^{\left\abs\hat U(2n_1+h)-n_2\right\abs^2\over 4\, \Im \!\hat T_1\Im \!\hat U_1}+\O(e^{-\hat c\, \tau_2\Im \hat T_1})\nonumber \\
&={\Im \hat T_1\over 2\tau_2\eta^2\bar\eta^2}\sum_{\tilde n_1, \tilde n_2}(-1)^{h\tilde n_1}\, e^{-{\pi\Im\!\hat T_1\over \tau_2 4\Im\!\hat U_1}\left\abs \tilde n_1+2\tilde n_2 \hat U_1\right\abs^2}+\O\big(e^{-\hat c\,\tau_2\Im \hat T_1}\big) ,
\end{align}
where $\hat c=\O(1)$ is positive and the second line is obtained by Poisson summation over $n_1$ and $n_2$. For $g=1$, the winding numbers cannot vanish, so that  
\be
\label{O2}
O_{2,2}^{(1)}\big[{}^1_h\big]\!_{(\hat T_1,\hat U_1)}=\O\big(e^{-\hat c\, \tau_2 \Im \hat T_1}\big) .
\ee

The light spectrum arising in Region (II) turns out to be : 

\vspace {0,3cm}

\noindent {\large \em Sector $O^{(1)}_{2,2}\big[{}^0_0\big]\!_{(\hat T_1,\hat U_1)}$}

\noindent This sector being self-dual, its massless spectrum is that derived in Sector $O^{(1)}_{2,2}\big[{}^0_0\big]\!_{(T_1,U_1)}$, which amounts to $n_{\rm B}$ bosonic and $n_{\rm F}$ fermionic degrees of freedom,  
\begin{align}
&n_{\rm B}\equiv d(\widehat{\rm Bosons}\big[{}^0_0\big])&&\!\!\!\!\!\!\!\!\!\!\!\!\!\!\!\!\!\!\!\!\!\!\!\!\!\!\!\!\!\!\!\!\!\!\!\!\!\!\!\!\!\!\!\!\!\!\!\!\!\!=8\times \big[244+d(G^{(2)})+d(G^{(3)})\big] 
 \nonumber \\
&n_{\rm F}\equiv d(\widehat{\rm Fermions}\big[{}^0_0\big])&&\!\!\!\!\!\!\!\!\!\!\!\!\!\!\!\!\!\!\!\!\!\!\!\!\!\!\!\!\!\!\!\!\!\!\!\!\!\!\!\!\!\!\!\!\!\!\!\!\!\!=8\times 256 \, . 
\end{align}
In  Regime (II), these degrees of freedom are accompanied by light towers of pure T-dual KK modes (pure winding modes for the original $1^{\rm st}$ 2-torus), whose momenta are $2n_1$ and $n_2$, as can be read in the $1^{\rm st}$ line of Eq. (\ref{O1}). Their masses are of order $M_{\rm s}/\sqrt{\Im \hat T_1/2}$.

\vspace {0,3cm}
\noindent {\large \em Sector $O^{(1)}_{2,2}\big[{}^0_1\big]\!_{(\hat T_1,\hat U_1)}$}

\noindent The fermionic sector $-O^{(0)}_{2,2}  O^{(1)}_{2,2}\big[{}^0_1\big]\!_{(\hat T_1,\hat U_1)}  O^{(2)}_{2,2} O^{(3)}_{2,2} C_8 \bar V_{16} \bar V'_{16}$ contains light towers of pure T-dual KK modes with momenta $2n_1+1$ and $m_2$. The former being nonzero, these states cannot be massless but their masses are light, of order $M_{\rm s}/\sqrt{\Im \hat T_1/2}$. Their degeneracy is 
 \begin{equation}
d(\widehat{\rm Fermions}\big[{}^0_1\big])= d(C_8)\,d(\bar V_{16})\, d(\bar V'_{16})=8\times 16\times 16=\underline{8\times 256}\, ,
\end{equation} 
which equals $n_{\rm F}$.

Note that no light bosonic state arises in Sector $O^{(1)}_{2,2}\big[{}^0_1\big]\!_{(\hat T_1,\hat U_1)}$, as can be seen from the right-moving characters $\bar V_{16} \bar C'_{16}+ \bar C_{16}\bar V'_{16}$, which start at the massive level, in units of $M_{\rm s}$. This shows that contrary to the large $\Im T_1$ limit with $U_1=\O(i)$, $\N=4$ supersymmetry is not recovered in the large $\Im \hat T_1$ limit  when $\hat U_1=\O(i)$. If the sss-model realizes a spontaneous breaking of supersymmetry implemented \via stringy Scherk-Schwarz compactification on the initial $1^{\rm st}$ 2-torus, from the T-dual picture, it realizes a compactification on the T-dual $\mbox{2-torus}$ of an initially non-supersymmetric model in 6 dimensions.  In fact, the dual KK mass scale $M_{\rm s}/\sqrt{\Im \hat T_1/2}$ is not a scale of supersymmetry breaking (spontaneous or not). The no-scale modulus \ie the spontaneous supersymmetry breaking scale is always $m_{3/2}$, which satisfies
\be
m^2_{3/2}={\abs U_1\abs^2 M^2_{\rm s}\over \Im T_1\, \Im U_1}= {\, \big\abs\hat T_1\big\abs^2\over 4} {M^2_{\rm s}\over \Im \hat T_1\, \Im \hat U_1}\gg M_{\rm s}^2\quad \mbox{in Regimes (II) and (III)}\, .
\ee 
In the limit $m^2_{3/2}\to +\infty$, degrees of freedom  decouple, leaving us with an sss-breaking of supersymmetry in six dimensions that is explicit.
 
The above remarks suggest that the vacuum energy may be large in Regime (II). To show this is true, we use Eqs (\ref{O1}) and (\ref{O2}) to write the effective potential in terms of dual moduli as
\begin{align}
\hat \V_{\mbox{\scriptsize 1-loop}}=&-{M_{\rm s}^4\over (2\pi)^4}\int_\F {d^2\tau\over 2\tau_2^2} \, O^{(0)}_{2,2} \, O^{(2)}_{2,2} \,O^{(3)}_{2,2}\, {\Im \hat T_1\over 2\tau_2\eta^2\bar\eta^2}\sum_{\tilde n_1,\tilde n_2}e^{-{\pi\Im\!\hat T_1\over \tau_2 4\Im\!\hat U_1}\left\abs \tilde n_1+2\tilde n_2 \hat U_1\right\abs^2}\; \times \nonumber \\
&\;\;\;\;\;\;\;\;\;\;\;\;\;\;\Big[\Big( V_8(\bar O_{16}  \bar O'_{16}+ \bar S_{16} \bar S'_{16} ) -S_8 (\bar O_{16}\bar S'_{16} + \bar S_{16} \bar O'_{16})\Big)\nonumber\\
&\; +(-1)^{\tilde n_1}\Big( O_{8} (\bar V_{16} \bar C'_{16}+ \bar C_{16}\bar V'_{16} ) -C_8 (\bar V_{16}\bar V'_{16} + \bar C_{16}\bar C'_{16} )\Big)\Big]\!+{\cal O}\Big(M_{\rm s}^4\, e^{-\hat c\sqrt{\Im \hat T_1}}\Big).
\end{align}
Contrary to the expression found  in  Regime (I) for large $\Im T_1$, the argument of the exponential in the $1^{\rm st}$ line, which is proportional to  $\abs \tilde n_1+2\tilde n_2 \hat U_1\abs^2$, can vanish. Actually, the contribution of the effective potential arising for $\tilde n_1=\tilde n_2=0$ grows linearly with  the dual volume $(2\pi)^2\Im \hat T_1/(2M_{\rm s}^2)$. This behavior is drastically different to that encountered in  Regime~(I), where the potential is exponentially suppressed in $\Im T_1$ (or scales like $(n_{\rm F}-n_{\rm B})m_{3/2}^4~$ if $n_{\rm F}\neq n_{\rm B}$) and vanishes in the limit where $\N=4$ supersymmetry is restored.  The remaining terms, with $(\tilde n_1,\tilde n_2)\neq(0,0)$, can be treated exactly as is done in  Regime (I) and mentioned in the introduction, in the paragraph above Eq. (\ref{v}). They yield light T-dual KK modes of masses $\O\Big(M_{\rm s}/\sqrt{\Im \hat T_1/2}\Big)$, whose contributions dominate over those arising from the remaining, super heavy string modes. Moreover, as follows from the $2^{\rm nd}$  line in Eq. (\ref{O1}), these towers of T-dual KK modes regularize the UV, in the sense that up to exponentially suppressed terms, the integral over the fundamental domain $\F$ can be extended to the upper half strip, $-{1\over 2}<\tau_1<{1\over 2}$, $\tau_2>0$, without introducing divergences.  In total, one finds 
\begin{align}
\hat\V_{\mbox{\scriptsize 1-loop}}&=C\,\Im \hat {T_1\over 2}+{n_{\rm F}-n_{\rm B}\over 16\pi^7}\, {M_{\rm s}^4\over (\Im \hat T_1)^2}\, {E_{(0,0)}(\hat U_1\abs 3,0)+E_{(1,0)}(\hat U_1\abs 3,0)\over 2}\nonumber \tesp\\
&+C'\Im \hat {T_1\over 2}+{n_{\rm F}\over 16\pi^7}\, {M_{\rm s}^4\over (\Im \hat T_1)^2}\, {E_{(0,0)}(\hat U_1\abs 3,0)-E_{(1,0)}(\hat U_1\abs 3,0)\over 2}+{\cal O}\Big(M_{\rm s}^4\, e^{-\hat c\sqrt{\Im \hat T_1}}\Big),
\end{align}
where the $1^{\rm st}$ and $2^{\rm nd}$ lines arise respectively  from the sectors $O_{2,2}^{(1)}\big[{}^0_0\big]\!_{(\hat T_1,\hat U_1)}$ and $O_{2,2}^{(1)}\big[{}^0_1\big]\!_{(\hat T_1,\hat U_1)}$,  while the quantities $C$ and $C'$ depend on the $2^{\rm nd}$ and $3^{\rm rd}$ 2-tori moduli only,
\begin{align}
C&=-{M_{\rm s}^4\over (2\pi)^4}\int_\F {d^2\tau\over 2\tau_2^4}  \,\Gamma_{2,2}{}_{(T_2,U_2)}\, \Gamma_{2,2}{}_{(T_3,U_3)}\!\left[ {V_8\over \eta^8}\, {\bar O_{16}  \bar O'_{16}+ \bar S_{16} \bar S'_{16} \over \bar \eta^8} -{S_8\over \eta^8} \, {\bar O_{16}\bar S'_{16} + \bar S_{16} \bar O'_{16}\over \bar \eta^8}\right],\nonumber \tesp\\
C'&=-{M_{\rm s}^4\over (2\pi)^4}\int_\F {d^2\tau\over 2\tau_2^4}  \,\Gamma_{2,2}{}_{(T_2,U_2)}\, \Gamma_{2,2}{}_{(T_3,U_3)}\!\left[ {O_8\over \eta^8}\,{\bar V_{16}  \bar C'_{16}+ \bar C_{16} \bar V'_{16} \over \bar \eta^8} -{C_8\over \eta^8} \,{\bar V_{16}\bar V'_{16} + \bar C_{16} \bar C'_{16}\over \bar \eta^8}\right].
\end{align}
The final expression of the effective potential in  Regime (II) can be simplified to 
\begin{align}
\label{pot}
\hat \V_{\mbox{\scriptsize 1-loop}}=\Big(C+C'\Big)&\Im \hat {T_1\over 2}+{M_{\rm s}^4\over 16\pi^7\, (\Im \hat T_1)^2}\; \times\nonumber\desp \\
&\left[\Big(n_{\rm F}-{n_{\rm B}\over 2}\Big)E_{(0,0)}(\hat U_1\abs 3,0)-{n_{\rm B}\over 2}E_{(1,0)}(\hat U_1\abs 3,0)\right]+{\cal O}\Big(M_{\rm s}^4\, e^{-\hat c\sqrt{\Im \hat T_1}}\Big).
\end{align}
Note that since $C+C'$ is nonzero, one obtains in the T-dual 2-torus decompactification limit
\be
\label{vac}
\int d^4x \, \hat \V_{\mbox{\scriptsize 1-loop}}\underset{\Im \hat T_1\to \infty}{\longrightarrow}\int d^4x \, \Im {\hat T_1\over 2}\, (C+C')=\int d^6x \, \hat \V^{\N_6=0}_{\mbox{\scriptsize 1-loop}}\, , 
\ee
where $\hat \V^{\N_6=0}_{\mbox{\scriptsize 1-loop}}~$ is the effective potential of the obtained non-supersymmetric six-dimensional theory,
\be
\hat \V^{\N_6=0}_{\mbox{\scriptsize 1-loop}}=-{M_{\rm s}^6\over (2\pi)^6}\int_\F {d^2\tau\over 2\tau_2^2}\, \hat Z_{\N_6=0}\, ,
\ee
which involves the associated partition function 
\begin{align}
 \hat Z_{\N_6=0}=O_{4,4}^{(0)}\; O_{2,2}^{(2)}\; O_{2,2}^{(3)}\, \Big(& V_8(\bar O_{16}  \bar O'_{16}+ \bar S_{16} \bar S'_{16} ) -S_8 (\bar O_{16}\bar S'_{16} + \bar S_{16} \bar O'_{16})\nonumber \\
 +\, &O_{8} (\bar V_{16} \bar C'_{16}+ \bar C_{16}\bar V'_{16} ) -C_8 (\bar V_{16}\bar V'_{16} + \bar C_{16}\bar C'_{16} )\Big) .
\end{align}
For instance, $C+C'$ can be evaluated numerically at $T_2=U_2=T_3=U_3=i$, which corresponds to the $G^{(2)}\times G^{(3)}=SU(2)^4$ enhanced symmetry point : $C+C'\simeq 0.468\, M_{\rm s}^4$. 

It is however important to stress that  the behavior of the sss-model derived  in Regimes (II) and (III) is actually formal. This is due to the fact that in these cases, the 1-loop correction to the classically vanishing vacuum energy density of the universe is very large, $\O(M_{\rm s}^6)$, as can be seen from the r.h.s. of Eq. (\ref{vac}). This fact may cast doubts on the validity of perturbation theory. Moreover, it is  expected that in the large T-dual $\mbox{2-torus}$ limit, the decompactification problem does arise. This should be the case since no $\N_6=2$ supersymmetry is recovered in six dimensions ($\N=4$ in four dimensions) and the towers of T-dual KK modes of masses $\O\Big(M_{\rm s}/\sqrt{\Im \hat T_1 /2}\Big)$ should yield large quantum corrections to the gauge thresholds, proportional to the volume $(2\pi)^2\Im \hat T_1/(2M_{\rm s}^2)$ \cite{solving,N=0thresh}. Finally, taking $n_{\rm F}\simeq n_{\rm B}$, which is satisfied for arbitrary $T_I,U_I$, $I\in\{2,3\}$, one can extremize the potential~(\ref{pot}) with respect to $\hat U_1$, which yields a solution $\hat U_1\simeq (1+i)/2$ modulo T-duality.  However, the latter is  a saddle point that destabilizes $\Im \hat U_1$ to larger and larger or lower and lower  values, which brings the theory out of  Regime (II). 


\subsection{The intermediate regime}
\label{InterRegime}

We proceed with the description of the behavior of the sss-model when no  modulus associated to the $1^{\rm st}$ 2-torus is large or small, \ie $T_1=\O(i)$, $U_1=\O(i)$. In this regime, $m_{3/2}=\O(M_{\rm s})$ and  the effective potential is not exponentially suppressed. Moreover, the generic massless states encountered in Sector $O^{(1)}_{2,2}\big[{}^0_0\big]\!_{(T_1,U_1)}$ are not accompanied anymore by light pure KK modes, the latter having masses of order $M_{\rm s}$. However, states with non-trivial momentum and winding numbers along the $1^{\rm st}$ 2-torus may be massless at special points in moduli space.

\vspace {0,3cm}
\noindent {\large \em Sector $O^{(1)}_{2,2}\big[{}^0_0\big]\!_{(T_1,U_1)}$}

\noindent Beside the generic massless bosons, additional ones in Sector $O^{(0)}_{2,2} O^{(1)}_{2,2}\big[{}^0_0\big]  O^{(2)}_{2,2} O^{(3)}_{2,2}   V_8 \bar O_{16} \bar O'_{16}$ become massless when ${1\over 2}\abs p_L^{(1)}\abs ^2={1\over 2}\abs p_R^{(1)}\abs^2-1=0$, thus increasing $n_{\rm B}$. For instance, taking $k_1=n_1=0$, these conditions are satisfied for $m_2=-n_2=\pm1$ when we sit on the codimension one submanifold of the moduli space that satisfies $T_1=-1/U_1$. These states are 2 gauge bosons and their Wilson lines along the internal space, 
\begin{equation}
\Delta n_{\rm B}\equiv d(\mbox{Extra Bosons}\big[{}^0_0\big])= d(V_8)\times d(O^{(1)}_{2,2}\big[{}^0_0\big])=\underline{8\times 2}\, ,
\end{equation}
which enhance the gauge group factor associated to the $1^{\rm st}$ 2-torus to $G^{(1)}=U(1)\times SU(2)$. On the contrary, $n_{\rm F}$ does not vary with $T_1, U_1$. 

\vspace {0,3cm}
\noindent {\large \em Sector $O^{(1)}_{2,2}\big[{}^1_1\big]\!_{(T_1,U_1)}$}

\noindent Other extra massless bosons arise in Sector $O^{(0)}_{2,2} O^{(1)}_{2,2}\big[{}^1_1\big]  O^{(2)}_{2,2} O^{(3)}_{2,2} O_8 \bar V_{16} \bar V'_{16}$ when ${1\over 2}\abs p_L^{(1)}\abs^2-{1\over 2}={1\over 2}\abs p_R^{(1)}\abs^2=0$. For instance, taking $m_2=n_2=0$, these conditions are satisfied for  $2k_1+1=2n_1+1=\pm 1$ when $T_1/2=-\bar U_1$. These modes are two scalars in the bi-fundamental representation of $SO(16)\times SO(16)'$, thus with multiplicity
 \begin{equation}
\Delta n_{\rm B}\equiv d(\mbox{Extra Bosons}\big[{}^1_1\big])= d(O^1_{2,2}\big[{}^1_1\big])\, d(\bar V_{16}) \, d(\bar V'_{16})=\underline{2\times 16\times 16} \, .  
\end{equation}

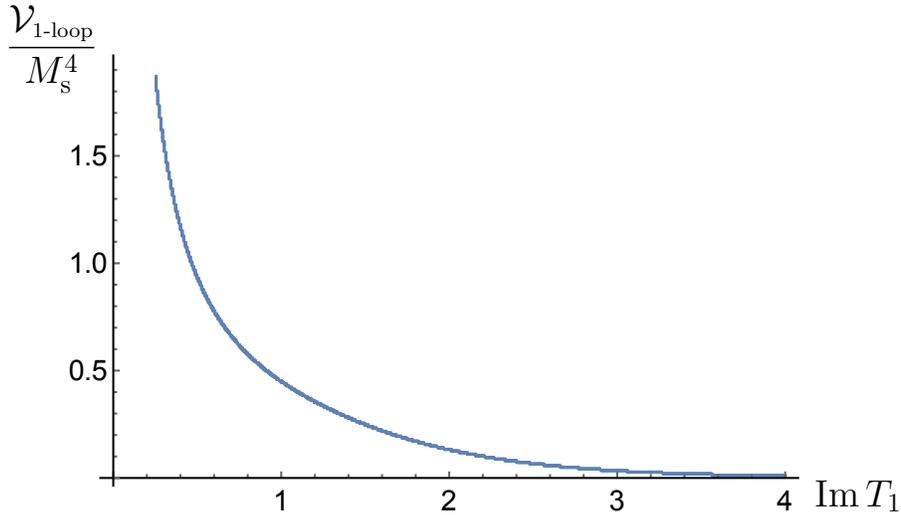
\begin{figure}[!t]
\begin{center}
\includegraphics[height=6.2cm]{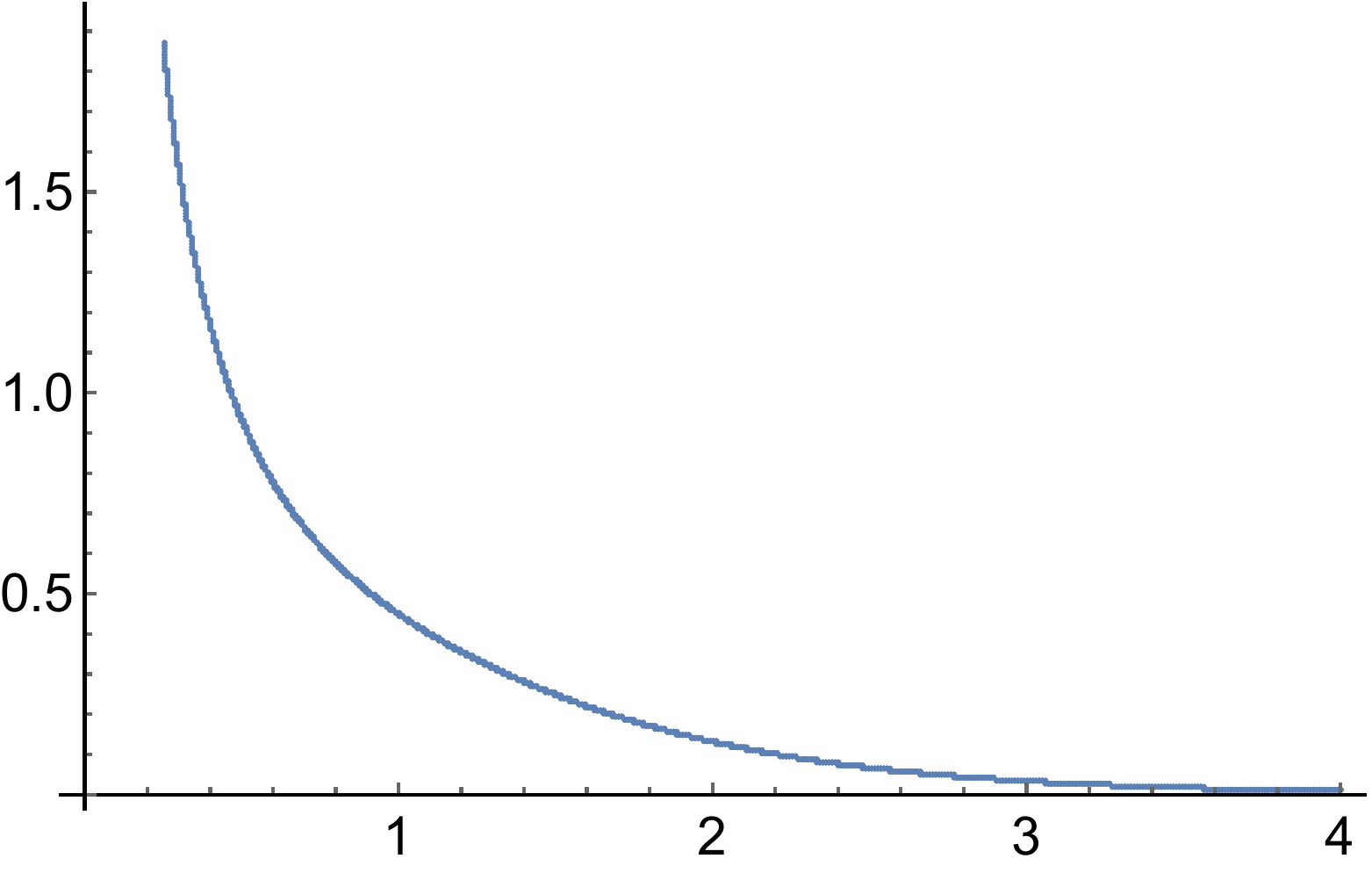}
\begin{picture}(6,5)
\put(1,5){\mbox{\large $\Im T_1$}}\put(-305,175){\mbox{\large $\displaystyle {\V_{\mbox{\scriptsize 1-loop}}\over M_{\rm s}^{4}}$}}
\end{picture}
\end{center}
\vspace{-0.5cm}
\caption{\footnotesize Effective potential of the sss-model as a function of $\Im T_1$, for $\Re T_1=0$, $U_1=T_2=U_2=T_3=U_3=i$. Regime (II), which corresponds to small $\Im T_1$,  is connected to the super no-scale Regime (I), where $\Im T_1$ is large.}
\label{fig1}
\end{figure}

Unlike the situation encountered in  Regimes (I)--(IV), no subset of string states, such as pure KK or winding modes, dominates the  expression (or part of it) of the effective potential. Moreover, the latter now depends on $\Re T_1$. Even if finding an explicit expression of $\V_{\mbox{\scriptsize 1-loop}}$ in the intermediate regime is a hard task, a numerical integration of the full partition function $Z_{\rm sss}$ can  always be  done over the fundamental domain $\F$. We choose to present the result as a function of $\Im T_1$ only, fixing $\Re T_1=0$ and $U_1=i$, while $T_2=U_2=T_3=U_3=i$. Generically, the gauge group is $G=G^{(1)}\times SU(2)^4\times SO(16)\times SO(16)'$, where $G^{(1)}=U(1)^2$. Fig. \ref{fig1} presents the curve  $\V_{\mbox{\scriptsize 1-loop}}$ as a function of $\Im T_1$ in these conditions. We see that the 1-loop effective potential is a positive and monotonically decreasing function, which connects Regime (II), where $m_{3/2}=M_{\rm s}/\sqrt{\Im T_1}\gg 1$, to the super no-scale Regime (I), where  $m_{3/2}\ll 1$.   
This behavior implies that  the term $e^{4\phi}\V_{\mbox{\scriptsize 1-loop}}$, which appears in the effective action in Einstein frame, creates a tadpole for the dilaton $\phi$ and imposes  the latter to  slide at  early cosmological times to the weak coupling regime. 

Our choice of $\Re T_1$ and $U_1$ is such that the curve passes through the lines $T_1=-1/U_1$ and $T_1/2=-\bar U_1$, when $\Im T_1=1$ and 2, respectively. However, no extremum occurs at these points. In Ref. \cite{GV}, it is shown in general that in  non-supersymmetric classical models, the integrated  partition function at arbitrary genus-$g$  admits extrema at all ``points of maximal enhanced symmetry". The latter are the loci in moduli space where the gauge group is enhanced, with no $U(1)$ factor left. In our case, since $G^{(1)}=U(1)\times SU(2)$ and  $G^{(1)}=U(1)^2$ at $\Im T_1=1$ and 2, there is no contradiction in not having extrema at these points. Fig.~\ref{fig1} shows that there exist initial conditions with $m_{3/2}$ of order $M_{\rm s}$, such that the super no-scale Regime (I) of Solution ($a$) is reached dynamically. However, as mentioned after Eq.~(\ref{modspace}), Wilson lines may also develop  expectation values in the intermediate regime, so that the theory may end with a distinct gauge group of equal rank,  or even suffer (for large deformations) from a classical tachyonic instability.   


\section{\bm $T^6$-moduli and Wilson lines deformations}
\label{deformations}

Once we have found a classical model that yields  an exponentially suppressed  effective potential at 1-loop, the question of the quantum stability of this background must be addressed. Actually, the worldsheet CFT admits marginal deformations, which from the spacetime point of view correspond to classical moduli. Since the 1-loop effective potential depends on these scalar deformations, the initial vacuum may be destabilized. In this section, we will study the response of the 1-loop effective potential to all   worldsheet small marginal deformations, in the super no-scale regime. As an example, we consider in details the case of Background~($a$) of the sss-model but the structure of the result remains valid in any generic $\N=4\to 0$ no-scale model \ie with $n_{\rm F}$ and $n_{\rm B}$ not necessary equal, and which is based in a gauge symmetry $U(1)^2\times \H$, where the rank of $\H$ is 20 and otherwise arbitrary.

\subsection{\bm Deformation of Background ($a$)}

The worldsheet operators we consider are $Y_{IJ}\partial X^I\bar \partial X^J$ and  $Y_{I\I}\partial X^I\bar \partial \bar \phi^\I$ for $I,J\in\{1,\dots,6\}$, $\I\in\{7,\dots,22\}$, where the $\bar \phi^\I$'s are the 16 extra right-moving compact bosons of the heterotic string.  In Background ($a$), we have initially $T_2=U_2=T_3=U_3=i$, the gauge group is   
\be
{\cal G}=U(1)^2\times SU(2)^4\times  SO(16)^2\, ,
\ee
and the partition function is given in Eq. (\ref{Zsss}), with $\Im T_1\gg 1$, $U_1=\O(i)$. Denoting $G_{IJ}$ and $B_{IJ}$ the initial internal metric and antisymmetric tensor, $6\times 6$ real $Y$'s are introduced to define their deformed counterparts as
\begin{align}
&(B'+G')_{\alpha\beta}=(B+G)_{\alpha\beta}+\sqrt{2}\, Y_{\alpha\beta}\, , \qquad\alpha,\beta\in\{1,2\}\, , \\
&(B'+G')_{ij}\;=\delta_{ij}+\sqrt{2}\, Y_{ij}\, , \,\quad \;\quad \qquad i,j\in\{3,4,5 ,6\}\, , \\
&(B'+G')_{\alpha i}\,=\sqrt{2}\, Y_{\alpha i}\, , \\
&(B'+G')_{i\alpha}\,=\sqrt{2}\, Y_{i\alpha}\, ,
\end{align}
while the  $6\times 16$ remaining ones, 
\be
Y_{\alpha \I}\, , \quad Y_{i \I}\, , \qquad \alpha\in\{1,2\}\, , \; i\in\{3,4,5,6\}\, , \; \I\in\{7,\dots,22\}\, , 
\ee
are the  Wilson lines  of $SO(16)^2$ along $T^6$.
Our goal is to determine which of the above  $6\times 22$ deformations  acquire at 1-loop positive squared masses or remain massless, while the leftover ones induce tachyonic instabilities. 
 
We first derive a general expression for the 1-loop effective potential, in the regime $\Im T_1\gg 1$, $U_1=\O(i)$. Let us consider the contribution to the 1-loop partition function arising from a single state $s$,
\be
\label{lm}
(-1)^F\, {1\over\tau_2}\, q^{{1\over 4}M_L^{\prime2}/M_{\rm s}^2}\, \bar q^{{1\over 4}M_R^{\prime2}/M_{\rm s}^2}=(-1)^F\, {1\over\tau_2}\,e^{-\pi\tau_2 M_L^{\prime2}/M_{\rm s}^2}\, \bar q^{{1\over 4}\left(M_R^{\prime2}-M_L^{\prime2}\right)/M_{\rm s}^2} , 
\ee
where $F$ is its fermion number. The left- and right-moving squared masses take the following form, where the ``primes" mean that the expressions refer to the deformed background \cite{421}, 
\be
\label{MLR}
M_L^{\prime2}= M_{\rm s}^2\Big[P_I'\, G^{\prime -1}_{IJ}P_J'+4\Big(N_L-{1\over 2}\Big)\Big] , \quad M_R^{\prime2}= M_{\rm s}^2\Big[\bar P_I'\, G^{\prime-1}_{IJ}\bar P_J'+Q'_\I Q'_\I+4(N_R-1)\Big] , 
\ee
where $N_L,N_R$ denote the oscillator numbers and we have defined 
\begin{align}
\label{PPQ}
P_I&=m_I+Y_{I\I}\, Q_\I+{1\over 2}Y_{I\I}\, Y_{J\I}\, n_J+(B'+G')_{IJ}\, n_J\, ,\nonumber \\
\bar P_I&=m_I+Y_{I\I}\,Q_\I+{1\over 2}Y_{I\I}\, Y_{J\I}\, n_J+(B'-G')_{IJ}\, n_J\, ,\nonumber \desp\\
Q'_\I&=Q'_\I+Y_{I\I}\, n_I\, .
\end{align}
In the above expressions, $P_I$ and $\bar P_I$ are generalized left- and right-moving momenta that depend on the $T^6$ momenta and winding numbers $m_I$ and $n_I$, while $Q_\I$, $\I\in\{7,\dots,22\}$, denote the components of a weight in a representation  of the gauge group realized by the extra right-moving $\bar \phi^\I$'s \cite{Lust-Theisen}. Physically, this weight is the charge vector under $SO(16)^2$ of the state $s$ and its squared length is an even integer.  An immediate consequence of the r.h.s. of Eq. (\ref{lm}) is that invariance under the modular translation $\tau\to \tau+1$ implies that $M_L^{\prime 2}-M_R^{\prime 2}=4L_s M_{\rm s}^2$, for some integer $L_s$. Therefore, $M_L^{\prime2}-M_R^{\prime2}$ must be invariant under the $6\times 22$ continuous deformations, a fact that is easily verified using Eqs (\ref{PPQ}), which yield $L_s=4m_In_I-2Q_\I Q_\I+4(N_L-{1\over 2})-4(N_R-1)$.

Next, we note that for small $Y$-deformations, the term (\ref{lm}) integrated over the fundamental domain $\F$ leads a contribution of order $e^{-\Im T_1}$ to the effective potential if $s$ has non-trivial winding numbers along either of the two $1^{\rm st}$ internal  directions, which are large.  Therefore, we concentrate on the dominant contributions, which arise from the pure momentum states (\ie with $2n_1+h=n_2=0$ in Eq. (\ref{pLR})). Choosing one of them, $s_0$, with vanishing momenta $m_1=m_2=0$ ($m_1\equiv 2k_1+g$ in Eq. (\ref{pLR})), let us  gather the contributions to $\V_{\mbox{\scriptsize 1-loop}}$ of the KK towers associated to $X^1,X^2$ and based on this state. In the initial Background $(a)$, one obtains
\begin{align}
\label{KKcontri}
-{M_{\rm s}^4\over (2\pi)^4}(-1)^{F_0}&\int_\F{d^2\tau\over 2\tau_2^3}\sum_{m_1,m_2}(-1)^{m_1}\, e^{-\pi\tau_2{\abs U_1 m_1-m_2\abs^2\over \Im T_1\Im U_1}}\, q^{{1\over 4}M_{0L}^2/M_{\rm s}^2} \, \bar q^{{1\over 4}M_{0R}^2/M_{\rm s}^2}\nonumber \\
&=-{M_{\rm s}^4\over (2\pi)^4}(-1)^{F_0}\int_\F{d^2\tau\over 2\tau_2^4} \, \Im T_1\sum_{\tilde m_1,\tilde m_2}e^{-{\pi\Im T_1\over \tau_2\Im U_1}\abs \tilde m_1+{1\over 2}+U_1 \tilde m_2\abs^2}\, q^{{1\over 4}M_{0L}^2/M_{\rm s}^2} \, \bar q^{{1\over 4}M_{0R}^2/M_{\rm s}^2}\, , 
\end{align}
where $F_0, M_{0L}, M_{0R}$ are the fermion number and left- or right-moving masses of $s_0$. The  insertion $(-1)^{m_1}$ in the l.h.s. arises from the fermion number $F=F_0+m_1$. It translates the fact that a mass splitting of order $1/\Im T_1$ exists between bosons and fermions, as follows from the spontaneous breaking of supersymmetry and can be seen in the partition function~(\ref{Zsss}). This phase $e^{i\pi m_1}$ yields in the r.h.s., which is obtained by Poisson summation, a ${1\over 2}$-shift of the integer $\tilde m_1$. This shift implies that the integral in Eq. (\ref{KKcontri}) can be extended to  the full upper half-strip, $-1<\tau_1\leq 1$, $\tau_2>0$, without introducing UV divergences, and that the result differs from that obtained by integrating over  $\F$ by terms of order $e^{-c\sqrt{\Im T_1}}$. 

When the $Y$-deformations are switched on, $M_{0L},M_{0R}$ and more importantly the KK mass are slightly modified. The latter is initially the degree 2 polynomial in $m_1,m_2$, which appears in the argument of the exponential function in the l.h.s. of Eq. (\ref{KKcontri}), (and becomes the expression in Eq. (\ref{defm})). However,  for small enough $Y$'s, the full expression after poisson summation is still integrable over the upper half-strip (see Eq. (\ref{Vie2})). It follows that the integration over $\tau_1$ is straightforward, implying that the surviving dominant contributions to $\V_{\mbox{\scriptsize 1-loop}}$ arise from KK states $s$ that are level-matched, $L_s=0$. Moreover, since the states with vanishing winding numbers along $X^1$ and $X^2$ satisfy
\be
4L_s M_{\rm s}^2=M_L^{\prime2}-M_R^{\prime2}=M_L^{2}-M_R^{2}=M_{0L}^{2}-M_{0R}^{2}\, ,
\ee 
which is independent of $m_1,m_2$, the whole towers of KK modes based on the level-matched states $s_0$ are level matched as well. Writing the associated contribution,
\be
-{M_{\rm s}^4\over (2\pi)^4}(-1)^{F_0} \,\Im T_1\sum_{\tilde m_1,\tilde m_2}\int_0^{+\infty}{d\tau_2\over 2\tau_2^4} \,e^{-{\pi\Im T_1\over \tau_2\Im U_1}\left[\abs \tilde m_1+{1\over 2}+U_1 \tilde m_2\abs^2+\O(Y)\right]}\, e^{-\pi\tau_2(M_{0L}^2+\O(Y))/M_{\rm s}^2}\, , 
\ee
and changing the dummy variable of integration $\tau_2$ into $x=\tau_2/\Im T_1$, we see that when the mass $M_{0L}$ of the state $s_0$ in the initial Background ($a$)  is not vanishing, the result is exponentially suppressed. Therefore, we obtain the general expression of the 1-loop effective potential
\be
\label{V}
\V_{\mbox{\scriptsize 1-loop}}=-{M_{\rm s}^4\over (2\pi)^4}\sum_{s_0=1}^{n_B+n_F}(-1)^{F_0}\int_0^{+\infty}{d\tau_2\over 2\tau_2^3}\sum_{m_1,m_2}(-1)^{m_1}\,e^{-\pi\tau_2 M_L^{\prime2}/M_{\rm s}^2}+{\cal O}\!\left(M_{\rm s}^4\, e^{-c\sqrt{\Im T_1}}\right)\!,
\ee
where the sum extends over the set of massless states present in the initial Background ($a$), and $M_L'$ is the mass of the associated KK mode with momenta $m_1,m_2$ along $X^1,X^2$, once the moduli deformations are switched on. 

To proceed, we resume the states $s_0$ of the sss-model in Background ($a$), which satisfy $M_{0L}^2=M_{0R}^2=0$. The first condition imposes $N_L={1\over 2}$,  the second yields $N_R=0$ or 1, and  we  recall that their quantum numbers along $X^1,X^2$ are $m_1=m_2=n_1=n_2=0$. 

\noindent - In Sector
\begin{align}
O^{(0)}_{2,2} &\Big(O_{2,2}^{(1)}\big[{}^0_0\big]V_8-O_{2,2}^{(1)}\big[{}^0_1\big]S_8\Big) O^{(2)}_{2,2}\,  O^{(3)}_{2,2} \, \bar O_{16} \bar O'_{16}\nonumber \\
&={1\over \tau_2\,\bar\eta^{24}}\Big(\Gamma_{2,2}\big[{}^{0,\, 0}_{0,\, 0} \big]\!_{(T_1,U_1)}8-\Gamma_{2,2}\big[{}^{0,\, 0}_{1,\, 0} \big]\!_{(T_1,U_1)}8\Big)\Gamma_{{\rm Adj}\, SU(2)^4}\, \Gamma_{{\rm Adj} \,SO(16)^2}+\O(q\bar q)\, ,
\end{align}
where $\Gamma_{{\rm Adj}\, SU(2)^4}$ and $\Gamma_{{\rm Adj} \,SO(16)^2}$ are the root lattices of $SU(2)^4$ and $SO(16)^2$, we find that : 

$\bullet$ At oscillator level $N_R=1$, 8 copies of 24 states with  $F_0=0$ arise from the factor $8/\bar \eta^{24}$. They are neutral with respect to the gauge group ${\cal G}$. $8\times 2$ realize the gravity sector, while the remaining  $8\times 22$ ones live in the Cartan subalgebra of ${\cal G}$. Their quantum numbers are 
\be
m_i=n_i=0\, , \; i\in\{3,4,5,6\}\, ,\qquad Q_\I=0\, , \; \I\in\{7,\dots,22\}\, .
\ee

$\bullet$ At oscillator level $N_R=0$, massless states with $F_0=0$ arise from the $SU(2)^4$ enhancement of the gauge symmetry. For any given $i\in \{3,4,5,6\}$ and $\epsilon\in\{-1,1\}$, there are 8 states with quantum numbers 
\be
m_i=-n_i=-\epsilon\, , \quad m_j=n_j=0\, , \; j\in\{3,4,5,6\},\, j\neq i\, ,\quad Q_\I=0\, , \; \I\in\{7,\dots,22\}\, .
\ee
Note that the generalized momentum $p_R^i={-1\over \sqrt{2}}(m_i/R_i-n_i R_i)$ of the compact direction $X^i$ of radius $R_i=1$ is  $p_R^i=\epsilon\sqrt{2}$, which is a root of squared length equal to 2 of $\Gamma_{{\rm Adj}\, SU(2)^4}$.

$\bullet$ Similarly, 8 copies of massless states with $F_0=0$ arise at oscillator level $N_R=0$ from the root lattice $\Gamma_{{\rm Adj} \,SO(16)^2}$. Their quantum numbers are 
\begin{align}
&m_i=n_i=0\, , \; i\in\{3,4,5,6\}\, ,\quad Q_\I\, , \; \I\in\{7,\dots, 22\} \, \; \where \nonumber\desp \\
&Q_\I=\left\{
\begin{array}{l}(\pm1,\pm1,0^6;0^8) \quad \mbox{or permutations between the entries $Q_7$ to $Q_{14}$,}\\\mbox{or}\\
(0^8;\pm1,\pm1,0^6)  \quad \mbox{or permutations between the entries $Q_{15}$ to $Q_{22}$.}
\end{array}
\right.
\end{align}
In the above formula, $0^k$ means $k$ consecutive null entries \cite{Lust-Theisen}. In total, there are $2\times 112$ such roots $Q_\I$ of squared lengths equal to  2. 

Altogether, we recover the $n_{\rm B}=8\times (24+4\times 2+2\times 112)=8\times 256$ bosonic massless states described in Sect. \ref{snsRegime}.

\noindent - In Sector 
\begin{align}
O^{(0)}_{2,2} &\Big(O_{2,2}^{(1)}\big[{}^0_1\big]V_8-O_{2,2}^{(1)}\big[{}^0_0\big]S_8\Big) O^{(2)}_{2,2}\,  O^{(3)}_{2,2} \, (\bar O_{16}\bar S'_{16} + \bar S_{16} \bar O'_{16})\nonumber \\
&={1\over \tau_2\,\bar \eta^{24}}\Big(\Gamma_{2,2}\big[{}^{0,\, 0}_{1,\, 0} \big]\!_{(T_1,U_1)}8-\Gamma_{2,2}\big[{}^{0,\, 0}_{0,\, 0} \big]\!_{(T_1,U_1)}8\Big) \Gamma_{{\rm Spin} \,SO(16)^2}+\O(q\bar q)\, ,
\end{align}
$\Gamma_{{\rm Spin} \,SO(16)^2}$ is the weight lattice of the spinorial representation of $SO(16)^2$. 8 copies of massless states with $F_0=1$ occur at oscillator number $N_R=0$ from this lattice. Their quantum numbers are 
\begin{align}
&m_i=n_i=0\, , \; i\in\{3,4,5,6\}\, ,\quad Q_\I\, , \; \I\in\{7,\dots, 22\}\, \; \where \nonumber \desp\\
&Q_\I=\left\{
\begin{array}{l}(\pm1,\pm1,\pm1,\pm1,\pm1,\pm1,\pm1,\pm1;0^8)\quad \mbox{with even number of $-1$'s,}\\\mbox{or}\\
(0^8;\pm1,\pm1,\pm1,\pm1,\pm1,\pm1,\pm1,\pm1)  \quad \mbox{with even number of $-1$'s.}
\end{array}
\right.
\end{align}
$Q_\I$ is actually one of the $2\times 128$ weights of squared lengths equal to 2 \cite{Lust-Theisen}.
As said in Sec.~\ref{snsRegime}, we have a total of $n_{\rm F} = 8\times(2\times128) = 8\times 256$ fermionic massless states. 

We are ready to compute the contribution to the effective potential (\ref{V}) that arises from the KK towers of states $s$ based on each of the  $n_B+n_F$ states $s_0$, which are initially massless in Background ($a$). The momenta, winding numbers and $SO(16)^2$ charges of each state $s$ are those of $s_0$, up to the momenta $m_1,m_2$ along the $X^1,X^2$, which are arbitrary. We first consider the non-Cartan states of $SU(2)^4$. For given $i\in\{3,4,5,6\}$ and $\epsilon\in\{-1,1\}$, the contribution of $s$ to the potential involves its  squared mass $M_L^{\prime 2}$ given in Eq. (\ref{MLR}), which is expressed in terms of  
\be
\label{Pi}
P'_\alpha= m_\alpha+\epsilon\Big(\sqrt{2}\, Y_{\alpha i}+Y_{\alpha \I}Y_{i\I}\Big)\, , \; \alpha\in\{1,2\}\, , \quad P'_j= \epsilon\Big(\sqrt{2}\, Y_{ji}+Y_{j \I}Y_{i\I}\Big)\, ,\; j\in\{3,4,5,6\}\, ,
\ee
and the inverse of the metric 
\be
G'_{IJ}=G_{IJ}+\sqrt{2}\, Y_{(IJ)} \qquad \where \qquad Y_{(IJ)}={1\over 2}(Y_{IJ}+Y_{JI})\, ,
\ee
which is 
\be
G^{\prime-1}_{IJ}=\left( \begin{array}{lr} \underline{G}^{-1}_{\alpha\beta}-\sqrt{2}\,  \underline{G}^{-1}_{\alpha \gamma}\, Y_{(\gamma\delta)}\, \underline{G}^{-1}_{\delta\beta}+\O(Y^2) & -\sqrt{2}\, \underline{G}^{-1}_{\alpha\gamma}\, Y_{(\gamma,k)}+\O(Y^2) \\ \\
-\sqrt{2}\, Y_{(j\gamma)}\, \underline{G}^{-1}_{\gamma,\beta}+\O(Y^2) & \delta_{jk}-\sqrt{2}\, Y_{(jk)}+\O(Y^2)\end{array}\right).
\ee
In the above equation, $\underline{G}^{-1}_{\alpha\beta}$ is the inverse of the $2\times 2$ matrix $\underline{G}_{\alpha\beta}=G_{\alpha\beta}$, $\alpha,\beta\in\{1,2\}$.  The contribution of the 8 copies of KK states associated to the $SU(2)^4$ root $i,\epsilon$ is
\begin{align}
\label{Vie}
\V^{i,\epsilon}_{\mbox{\scriptsize 1-loop}} = -{8\,M_{\rm s}^4\over 2(2\pi)^4}\int_0^{+\infty}{d\tau_2\over \tau_2^3}&\sum_{m_1,m_2}(-1)^{m_1} \, e^{-\pi \tau_2 (m_\alpha+\xi_\alpha)G^{\prime -1}_{\alpha\beta}(m_\beta+\xi_\beta)} \; \times\nonumber\\
&\Big[1-2\pi\tau_2 \, m_\alpha  \sqrt{2}\, \underline{G}^{-1}_{\alpha\gamma}\, Y_{(\gamma j)}\, \epsilon\sqrt{2}\; Y_{ji}+\O(Y^3)\Big]\, \times \nonumber \\
&\Big[1-\pi\tau_2\big(Y_{ji}\, \epsilon\sqrt{2}\big)^2+\O(Y^3)\Big]+{\cal O}\!\left(M_{\rm s}^4\, e^{-c\sqrt{\Im T_1}}\right)\desp,
\end{align}
where $\xi_\alpha=\epsilon\Big(\sqrt{2}\, Y_{\alpha i}+Y_{\alpha \I}Y_{i\I}\Big)$ and we have expanded at second order in $Y$'s the $e^{-\pi\tau_2 2 P_\alpha'\, G^{\prime -1}_{\alpha j}P_j'}$ and $e^{-\pi\tau_2 P_j'\, G^{\prime -1}_{jk}P'_k}$ contributions in the integrand. However, the $2^{\rm nd}$ line in Eq. (\ref{Vie}) can be omitted, since its linear term in $m_\alpha$ must be dressed, at the order we are interested in,  by $e^{-\pi \tau_2 m_\alpha G^{-1}_{\alpha\beta}m_\beta}$ coming from the $1^{\rm st}$ line, and we sum over $m_1,m_2$. Recalling the definition of the component $B'_{21}$ of the deformed antisymmetric tensor and choosing the $2\times 2$ matrix $\underline{G}^{\prime-1}_{\alpha\beta}$ as follows, 
\be
B_{21}'=B_{21}+{1\over \sqrt{2}}(Y_{21}-Y_{12})\, , \qquad \quad \underline{G}^{\prime-1}_{\alpha\beta}=G^{\prime -1}_{\alpha\beta}\, ,\quad \alpha,\beta\in\{1,2\}\, , 
\ee
we can use the inverse matrix\footnote{Note that $\underline{G}'_{\alpha\beta}$ differs from $G_{\alpha\beta}'$ at quadratic order in $Y$'s.} $\underline{G}'_{\alpha\beta}$ to define deformed moduli  
\be
T'_1=i\sqrt{\underline{G}'_{11}\underline{G}'_{22}-\underline{G}_{12}^{\prime2}}+B'_{21} \, , \qquad \quad  U'_1={i\sqrt{\underline{G}'_{11}\underline{G}'_{2,2}-\underline{G}_{12}^{\prime 2}}+\underline{G}'_{21}\over \underline{G}'_{11}}\, ,
\ee
in terms of which we have 
\be
\label{defm}
(m_\alpha+\xi_\alpha)\, G^{\prime -1}_{\alpha\beta}\, (m_\beta+\xi_\beta)={\abs U_1' (m_1+\xi_1)-(m_2+\xi_2)\abs^2\over \Im T_1' \, \Im U_1'}\, .
\ee
A Poisson summation on $m_1,m_2$ in $\V^{i,\epsilon}_{\mbox{\scriptsize 1-loop}}$ then leads 
\begin{align}
\label{Vie2}
\V^{i,\epsilon}_{\mbox{\scriptsize 1-loop}} = -{8\,M_{\rm s}^4\over 2(2\pi)^4}\int_0^{+\infty}{d\tau_2\over \tau_2^3}&\, {\Im T_1'\over \tau_2}\sum_{\tilde m_1,\tilde m_2}e^{-{\pi \Im T_1'\over \tau_2 \Im U_1'}\abs\tilde m_1+{1\over 2}+U_1'\tilde m_2\abs^2}\, e^{2i \pi {\Re [(\tilde m_1+{1\over 2}+U_1' \tilde m_2)\bar \xi]\over \Im U_1'}} \nonumber\\
&\Big[1-\pi\tau_2\big(Y_{ji}\, \epsilon\sqrt{2}\big)^2+\O(Y^3)\Big]+{\cal O}\!\left(M_{\rm s}^4\, e^{-c\sqrt{\Im T_1}}\right),
\end{align}
where $\xi=U_1 \xi_1-\xi_2$. Expanding the phase in $\xi$ or $\bar \xi$ and integrating over $\tau_2$, one obtains the final contribution,
\begin{align}
\label{Vi3}
\V^{i,\epsilon}_{\mbox{\scriptsize 1-loop}}=&-{8\over 16\pi^7}\,{ M_{\rm s}^4\over (\Im T'_1)^2}\, E_{(1,0)}(U'_1\abs 3,0)
+{8\over 16\pi^5}\,{ M_{\rm s}^4\over \Im T_1}\, E_{(1,0)}(U_1\abs 2,0)\; \times\desp \\
&\; \quad \qquad\qquad  {1\over 2}\!\left(\sum_{j=3}^6\big(Y_{ji}\,\epsilon\sqrt{2}\big)^2+2\big\abs Y_i\, \epsilon\sqrt{2}\big\abs^2-\rho\big(Y_i\, \epsilon\sqrt{2}\big)^2-\bar\rho\big(\bar Y_i\,  \epsilon\sqrt{2}\big)^2\right)\nonumber\tesp\\
&\hspace{6.9cm}+\O(M_{\rm s}^4\, Y^3)+{\cal O}\!\left(M_{\rm s}^4\, e^{-c\sqrt{\Im T_1}}\right)  ,
\end{align}
where we have  redefined complex moduli as 
\be
\label{Yi}
Y_i={U_1 Y_{1i}-Y_{2i}\over \sqrt{\Im T_1\Im U_1}}\, ,\qquad i\in\{3,4,5,6\}\, , 
\ee 
and introduced the dressing coefficient
\be
\rho={E_{(1,0)}(U_1\abs 2,1)\over E_{(1,0)}(U_1\abs 2,0)}\, .
\ee
In Eq. (\ref{Vi3}), the scalars $Y_{ji}$ and $Y_i$, for $j\in\{3,4,5,6\}$, are actually the Wilson lines of the $i^{\rm th}$ $SU(2)$ factor along $T^6$, weighted by the associated root $\epsilon\sqrt{2}$. 

We proceed with the contribution $\V^Q_{\mbox{\scriptsize 1-loop}}$ of the effective potential that arises from  the KK modes $s$ based on the state $s_0$ of  right-moving charge $Q_\I$, which is either a root of $\Gamma_{{\rm Adj} \,SO(16)^2}$ or  a weight of $\Gamma_{{\rm Spin} \,SO(16)^2}$ whose length squared equals 2. The novelty is that the former have $F_0=0$, while  the latter have $F_0=1$. For such a mode $s$, we have
\be
\label{Paj}
P'_\alpha= m_\alpha+Y_{\alpha \I}\, Q_\I\, , \quad \alpha\in\{1,2\}\, , \qquad \quad P'_j= Y_{j\I}\, Q_\I\, ,\quad j\in\{3,4,5,6\}\, .
\ee
Comparing with Eq. (\ref{Pi}), we see that at second order in $Y$'s, the 8 copies of KK modes yield a contribution identical to  $\V^{i,\epsilon}_{\mbox{\scriptsize 1-loop}}$, up to the overall dressing $(-1)^{F_0}$ and the exchanges 
\be
Y_{\alpha i}\, \epsilon \sqrt{2}\longrightarrow Y_{\alpha \I}\, Q_\I\, , \quad \alpha\in\{1,2\}\,,\qquad\quad  Y_{ji}\, \epsilon\sqrt{2}\longrightarrow  Y_{j\I}\, Q_\I\, ,\quad j\in\{3,4,5, 6\}\,.
\ee
Thus,  we immediately conclude that 
\begin{align}
\label{Vi3'}
\V^Q_{\mbox{\scriptsize 1-loop}}=&-{(-1)^{F_0}\, 8\over 16\pi^7}\,{ M_{\rm s}^4\over (\Im T'_1)^2}\, E_{(1,0)}(U'_1\abs 3,0)
+{(-1)^{F_0}\, 8\over 16\pi^5}\,{ M_{\rm s}^4\over \Im T_1}\, E_{(1,0)}(U_1\abs 2,0)\; \times\desp\nonumber  \\
&\; {1\over 2}\!\left(\sum_{j=3}^6\Big(\sum_{\I=7}^{22}Y_{j\I}\,Q_\I\Big)^2+2\Big\abs \sum_{\I=7}^{22}Y_\I\, Q_\I\Big\abs^2-\rho\Big(\sum_{\I=7}^{22}Y_\I\, Q_\I\Big)^2-\bar\rho\Big(\sum_{\I=7}^{22}\bar Y_\I\, Q_\I \Big)^2\right)\nonumber\tesp\\
&\hspace{7.3cm}+\O(M_{\rm s}^4\, Y^3)+{\cal O}\!\left(M_{\rm s}^4\, e^{-c\sqrt{\Im T_1}}\right)  ,
\end{align}
where $Y_{j\I}$, $j\in\{3,4,5,6\}$, and  
\be
Y_\I={U_1 Y_{1\I}-Y_{2\I}\over \sqrt{\Im T_1\Im U_1}}\, ,\qquad \I\in\{7,\dots,22\}\, , 
\ee 
are the Wilson lines of $SO(16)^2$ along $T^6$.

Finally, we consider the $8\times 24$ KK towers of states that are neutral with respect to the gauge group. In this case, $F_0=0$ and $P'_\alpha, P'_j$ are like those of Eq. (\ref{Paj}), with $Q_\I=0$. Therefore, the effective potential contribution $\V^e_{\mbox{\scriptsize 1-loop}}$, for $e\in\{1,\dots,24\}$, which arises from the 8 copies of such states, is
\be
\label{V4}
\V^e_{\mbox{\scriptsize 1-loop}}=-{8\over 16\pi^7}\,{ M_{\rm s}^4\over (\Im T'_1)^2}\, E_{(1,0)}(U'_1\abs 3,0) 
+{\cal O}\!\left(M_{\rm s}^4\, e^{-c\sqrt{\Im T_1}}\right)  .
\ee

In order to combine all contributions to the effective potential we have computed, we note that 
\begin{align}
&{1\over 2}\sum_{Q\in\!\!\!\!
\underset{\mbox{\scriptsize of $SO(16)^2$}}{\mbox{\scriptsize Adjoint}}}
\sum_{\I=7}^{22}A_\I Q_\I\, \sum_{\J=7}^{22}B_\J Q_\J=C({\cal A}_{SO(16)})\sum_{\I=7}^{22}A_\I B_\I\, ,  &&\mbox{for}& \!\!C({\cal A}_{SO(16)})=14\, , \nonumber \\
&{1\over 2}\sum_{Q\in\!\!\!
\underset{\mbox{\scriptsize of $SO(16)^2$}}{\mbox{\scriptsize Spinorial}}}
\sum_{\I=7}^{22}A_\I Q_\I\, \sum_{\J=7}^{22}B_\J Q_\J=C({\cal S}_{SO(16)})\sum_{\I=7}^{22}A_\I B_\I\, ,  &&\mbox{for}&\!\!C({\cal S}_{SO(16)})=16\, ,
\end{align}
where $C({\rm \cal R}_G)\delta^{ab}={\rm tr}(T^aT^b)$ and $T^a$, $a\in\{1,\dots,\dim G\}$, are the  generators in the representation ${\cal R}_G$ of a gauge group $G$. Given that, summing over the $4\times 2$ roots $i,\epsilon$ of $SU(2)^4$, the $2\times (112+128)$ charges $Q$ of $SO(16)^2$ and the 24 sets of neutral KK towers, one obtains  the final result,
 \begin{align}
\label{Vtot}
\V_{\mbox{\scriptsize 1-loop}}= & \; {n_{\rm F}-n_{\rm B}\over 16\pi^7}\,{M_{\rm s}^4\over (\Im T'_1)^2}\, E_{(1,0)}(U'_1\abs 3,0) \nonumber\desp \\
&\, -{3\over 16\pi^5}\,{ M_{\rm s}^4\over \Im T_1}\, E_{(1,0)}(U_1\abs 2,0)\Bigg(b_{SU(2)}\sum_{i=3}^6\!\bigg[\sum_{j=3}^6(Y_{ji})^2+2\abs Y_i\abs^2-\rho(Y_i)^2-\bar\rho(\bar Y_i)^2\bigg]\nonumber\tesp \\
&\hspace{4.1cm}+b_{SO(16)}\sum_{\I=7}^{22}\!\bigg[\sum_{j=3}^6(Y_{j\I})^2+2\abs Y_\I\abs^2-\rho(Y_\I)^2-\bar\rho(\bar Y_\I)^2\bigg]\Bigg)\nonumber\tesp \\
&\hspace{7.3cm}+\O(M_{\rm s}^4\, Y^3)+{\cal O}\!\left(M_{\rm s}^4\, e^{-c\sqrt{\Im T_1}}\right)  .
\end{align}
In this expression, $b_{SU(2)}$ and $b_{SO(16)}$ are the $\beta$-function coefficients of each $SU(2)$ and $SO(16)$ factors, 
\begin{align}
b_{SU(2)}&=\left(-{11\over 3}+6\times {1\over 6}\right)\!C(\A_{SU(2)})=-{8\over 3}\,2\, , \nonumber\\
b_{SO(16)}&=\left(-{11\over 3}+6\times {1\over 6}\right)\!C(\A_{SO(16)})+4\times {2\over 3}\,C(\S_{SO(16)})\nonumber \\
&= -{8\over 3}\big(C(\A_{SO(16)})-C(\S_{SO(16)})\big)=-{8\over 3}\,(-2)\, ,
\end{align}
which are obtained using the following contributions of massless degrees of freedom in the representation ${\cal R}_G$ of $G$,
\be
 \label{coeff}
b_G^{\mbox{\scriptsize gauge boson}} =-{11\over 3} \, C({{\cal R}_G})\, , \quad   b_G^{\mbox{\scriptsize real scalar}} ={1\over 6} \,  C({{\cal R}_G})\, , \quad b_G^{\mbox{\scriptsize Majorana fermion}} = {2\over 3}\, C({{\cal R}_G})\, .
\ee

Note that in the derivation of Eq. (\ref{Vtot}), the fact that $n_{\rm F}=n_{\rm B}$ in the sss-model plays no role. Thus, the above structure of the effective potential in terms of arbitrary $n_{\rm F}$, $n_{\rm B}$ and $\beta$-function coefficients associated to the simple gauge group factors is valid for arbitrary no-scale model realizing the $\N=4\to 0$ spontaneous breaking. 
In such a generic model, with $n_{\rm F}\neq n_{\rm B}$, the dominant term appearing in the  $1^{\rm st}$ line in Eq. (\ref{Vtot}) is proportional to $m_{3/2}^{\prime 4}$, where  $m_{3/2}'$ is the deformed gravitino mass,
\begin{equation}
\label{m32'}
m^{\prime 2}_{3/2}={\abs U'_1\abs^2 M^2_{\rm s}\over \Im T'_1\, \Im U'_1}\, .
\end{equation} 
 Observe  that since the moduli $Y_{(\alpha i)}$, $\alpha\in\{1,2\}$, $i\in\{3,4,5,6\}$,  are switched on, the  $T^2\times T^4$ factorized form of the internal space of the initial Background ($a$) is broken, implying $m_{3/2}'$ to depend on the whole metric of $T^6$.\footnote{The gravitino mass $m_{3/2}'$ involves $T_1',U_1'$ \ie $B_{21}$ and $\underline{G}'_{\alpha\beta}$ only, but the latter is the inverse $2\times 2$ matrix of $\underline{G}^{\prime -1}_{\alpha\beta}= \underline{G}^{-1}_{\alpha\beta}-\sqrt{2}\,  \underline{G}^{-1}_{\alpha \gamma}\, Y_{(\gamma\delta)}\, \underline{G}^{-1}_{\delta\beta}+2\,  \underline{G}^{-1}_{\alpha\gamma}\, Y_{(\gamma\delta)}\,  \underline{G}^{-1}_{\delta\rho}\, Y_{(\rho\sigma)}\,  \underline{G}^{-1}_{\sigma\beta}+ 2\, \underline{G}^{-1}_{\alpha\gamma}\, Y_{(\gamma j)}\, Y_{(j\delta)}\,  \underline{G}^{-1}_{\delta\beta}+\O(Y^3)$.} Clearly, the stability of an initial no-scale model background requires the term $m_{3/2}^{\prime 4}$ to be absent, which is nothing but the super no-scale condition $n_{\rm F}=n_{\rm B}$.
If this is satisfied, we are left with the $2^{\rm nd}$  and $3^{\rm rd}$ lines in Eq. (\ref{Vtot}), which are proportional to $m_{3/2}^2 M_{\rm s}^2$. The eigenvalues of  the squared mass matrices of the dimensionful scalars $Y_i M_{\rm s}$, $i\in\{3,4,5,6\}$ and $Y_\I M_{\rm s}$, $\I\in\{7,\dots,22\}$ are
\be
\label{inst}
-{3b_G\over 16\pi^5}\,{ M_{\rm s}^2\over \Im T_1}\, E_{(1,0)}(U_1\abs 2,0)\big(1\pm \abs\rho\abs\big)\, ,\quad \mbox{for $G=SU(2)$ or $SO(16)$} \,,
\ee
which are proportional to $m_{3/2}^2$, as expected for moduli not involved in the supersymmetry breaking \cite{FKZ}.
Since $\abs\rho(U_1)\abs<1$, Eq.~(\ref{inst}) leads to the conclusion  that {\em any simple gauge group factor that is neither asymptotically free nor conformal, i.e. with $b_G>0$, yields to local instabilities.} 

In the sss-super no-scale model we consider here, the $SU(2)^4$ Wilson lines $Y_{ji}$ and $Y_i$, $j,i\in\{3,4,5,6\}$, are attracted dynamically to the origin $Y_{ji}=Y_i=0$, while the $SO(16)^2$ ones $Y_{j\I}$ and $Y_\I$, $j\in\{3,4,5,6\}$, $\I\in\{7,\dots,22\}$, condense. Due to the periodicity properties of the Wilson lines, this instability is only local and some of the $Y_{j\I}$'s and/or $Y_\I$'s are expected to develop large but finite expectation values. Note that since we started with a vanishing effective potential in the super no-scale Background ($a$), these instabilities imply that $\V_{\mbox{\scriptsize 1-loop}}$ becomes negative. We should  reach another no-scale model, with new numbers of massless fermions and bosons satisfying $n'_{\rm F}<n'_{\rm B}$, and without non-asymptotically free gauge group factors. At this stage, the model would still be in the regime $m'_{3/2}\ll M_{\rm s}$, which guaranties no tachyonic instability may arise. However, the scaling of the effective potential now being like $-m_{3/2}^{\prime 4}$, the gravitino mass would be dynamically attracted to larger values. Once it  reaches the order of magnitude of the string scale, several scenarios may occur : 

$\bullet$ A tachyon may arise at tree level, thus inducing a severe Hagedorn-like instability.   

$\bullet$  $m_{3/2}$ may be stabilized at a (local) minimum, thus yielding an anti-de Sitter vacuum, where a restoration of supersymmetry may or may not occur.  

$\bullet$  $m_{3/2}$ may continue increasing, with runaway behavior. The model would lead (after T-duality) to an anti-de Sitter  theory in higher dimensions, explicitly non-supersymmetric.   


\subsection{Lifting the instabilities}
{\label{MarginalDeformations}}

In the previous sub-section we have shown the existence of two different types of instabilities. The first ones, arise in the no-scale models having  $n_{\rm F}\neq n_{\rm B}$, which are due to the non-vanishing of  $\V_{\mbox{\scriptsize 1-loop}}$.  Actually, the vanishing of the effective potential is required by  the dilaton and no-scale modulus stationary condition; namely  the absence of dilaton and no-scale modulus tadpoles. The second ones are tachyonic instabilities that arise in all no-scale models having positive $\beta$-function coefficients. 
Therefore, it would be relevant to look for super no-scale models without non-asympto\-ti\-cally free gauge group factors. Possibly, one could consider no-scale models with $n_{\rm F}> n_{\rm B}$, and switch on discrete Wilson lines of order 1 in order to break the non-asymptotically free gauge group factors to products of  asymptotically free  and/or conformal subgroups. 

Another approach is to consider the super no-scale models at finite temperature $T$. Note that this point of view can be relevant when the models are used in cosmological scenarios. At finite $T$, the effective potential is nothing but the quantum free energy and all squared masses are shifted by $T^2$ \cite{CosmologicalTerm}. Thus, as long as $T^2$ is greater than $m_{3/2}^2$, the tachyonic instabilities arising from positive $\beta$-function coefficients are lifted. For instance, Background~($a$) of the sss-model is stable during early stages of the cosmological evolution, when $T$ is high. As the Universe grows and the temperature drops,  the breaking of $SO(16)\times SO(16)'$ occurs when $T^2$ crosses $m_{3/2}^2$ and becomes lower. It would be interesting to  investigate this phase transition in a dynamical cosmological framework where all moduli fields, including the dilaton and the no-scale modulus, evolve with the temperature. 

Another way to bypass the tachyonic instabilities occurring at 1-loop in super no-scale models may be to impose correlations among  deformations, in order to preserve those which respect at the quantum level the flatness condition $\V_{\mbox{\scriptsize 1-loop}}=0$. In the case of Background~($a$), since $-b_{SU(2)}=b_{SO(16)}$, ideally the constraint 
 \be
 \label{flatness} 
 Y^2=H^2   
 \ee
 may be implemented, where $Y$ is the total ``attractive" Wilson line deformation associated to $SU(2)^4$, while $H$ is the total ``repulsive" one,  associated to $SO(16)^2$,
 \begin{align}
Y^2&=  \sum_{i=3}^6\!\bigg[\sum_{j=3}^6(Y_{ji})^2+2\abs Y_i\abs^2-\rho(Y_i)^2-\bar\rho(\bar Y_i)^2\bigg],\nonumber \\
H^2&=\sum_{\I=7}^{22}\!\bigg[\sum_{j=3}^6(Y_{j\I})^2+2\abs Y_\I\abs^2-\rho(Y_\I)^2-\bar\rho(\bar Y_\I)^2\bigg].
 \end{align}
 Differently  stated, one would demand  the negative energy density created by any breaking  of  $SO(16)^2$ to be compensated by the positive one, generated by a breaking of $SU(2)^4$. It may be relevant to investigate this possibility by implementing additional orbifold actions.  
 

\section{\bm $\N=2\to 0$ and $\N=1\to 0$ super no-scale models}
{\label{descendant}}

In the super no-scale models presented so far, with exponentially suppressed vacuum energies at the 1-loop quantum level, $\N=4$ supersymmetry is spontaneously broken to $\N=0$. It is then legitimate to look for less symmetric  super no-scale theories, realizing either an  
$\N=2 \to 0$ or $\N=1\to 0$ spontaneous breaking. For this purpose, one may consider no-scale parent theories describing an $\N=4\to 0$ breaking, and implement $\Z_2$ or $\Z_2\times \Z_2$ orbifold actions that yield descendent models satisfying the super no-scale property. However, as was  shown in Ref. \cite{N=0thresh},  if no precautions are taken in the choice of  orbifold actions, the $\N=2$ sectors of these models lead generically to gauge coupling threshold corrections \cite{thresholds,universality} proportional to the large internal volume \cite{solving}. In this case,  a fine tuning of the string coupling $g_{\rm s}$ is required  to cancel the 1-loop threshold corrections of the gauge couplings of  the  asymptotically free gauge group factors. In the following, we present a simple strategy that yields $\N=2\to 0$ or $\N=1\to 0$ super no-scale models, while evading the above mentioned  ``decompactification problem". 


\subsection{\bm Chains of $\N=4,2,1\to 0$ super no-scale models} 
\label{chain}
Our goal is to derive a class of $\N=2\to 0$ and $\N=1\to 0$ super no-scale models from parent ones that realize the $\N=4\to 0$ breaking. The next subsection will describe the gauge threshold corrections arising in this case. To begin, we consider any  $\N=4$ heterotic no-scale vacuum obtained by  ``moduli-deformed fermionic construction" \cite{fermionic, N=0thresh}. Let us implement a $\Z_2$ or $\Z_2\times \Z_2$ orbifold action where at least one of the $\Z_2$'s is freely acting and thus realizes a {\em spontaneous} $\N=4\to 2$ breaking. The resulting vacuum is $\N=2$ or $\N=1$ supersymmetric, which is further spontaneously broken to $\N=0$ by a stringy Scherk-Schwarz mechanism \cite{SSstring} realized along the $1^{\rm st}$ internal 2-torus. The latter is chosen to be large, for the supersymmetry breaking scale to be small, $m_{3/2}^2\propto M_{\rm s}^2/ \Im T_1$. In total, the model describes the $\N=4\to 2\to 0$ or $\N=2\to 1\to 0$ pattern of supersymmetry breaking. To be more specific, we request  the following \cite{N=0thresh} :

 $\bullet$ The generator of the free $\Z_2$ action, denoted as $\Z^{\rm free}_2$, twists the coordinates of the $2^{\rm nd}$ and $3^{\rm rd}$ 2-tori,  and shifts at least one of the coordinates of the $1^{\rm st}$ 2-torus, \eg
\be
\Z_2^{\rm free}\; : (X^1,X^2,X^3,X^4,X^5,X^6)\longrightarrow  (X^1,X^2+{1/ 2},-X^3,-X^4,-X^5,-X^6)\, .
\ee 

$\bullet$  In the $\Z^{\rm free}_2\times \Z_2$ case, there  is no restriction on  the second $\Z_2$. However, in most cases, its generator as well as the product of the latter with the generator of $\Z^{\rm free}$ have fixed points.  If this happens, in order  not to  induce large threshold corrections to the gauge couplings, we impose the $2^{\rm nd}$ and  $3^{\rm rd}$ 2-tori moduli $T_2,U_2$ and   $T_3,U_3$  not to be far from$~i$. For instance, they can sit at extended symmetry points. 

$\bullet$  The stringy Scherk-Schwarz mechanism responsible for the final supersymmetry breaking is realized as a ${1\over 2}$-shift  along  the $1^{\rm st}$ 2-torus, say $X^1$, coupled to one of the $R$-symmetry charges, such as the helicity $a$. 

Once the above restrictions are satisfied and $m_{3/2}\ll M_{\rm s}$, the effective potential of the $\Z_2^{\rm free}$  and $\Z^{\rm free}_2\times \Z_2$ models turn out to be  ${1\over 2}$ and ${1\over4}$ of that of the  ``parent" $\N=4\to 0$ theory,  up to exponentially suppressed contributions \cite{N=0thresh},
\begin{align}
&\V_{\mbox{\scriptsize 1-loop}}\Big\abs_{\N=4\to 2 \to 0}={1\over 2}\,\V_{\mbox{\scriptsize 1-loop}}\Big\abs_{\N=4 \to 0}+\O\!\left(M_{\rm s}^4\, e^{-cM_{\rm s}/ m_{3/2}}\right) ,\nonumber \phantom{\underset{\hat \abs}{\Big\abs}}  \\
&\V_{\mbox{\scriptsize 1-loop}}\Big\abs_{\N=2\to 1 \to 0}={1\over 4}\, \V_{\mbox{\scriptsize 1-loop}}\Big\abs_{\N=4 \to 0}+\O\!\left(M_{\rm s}^4\, e^{-cM_{\rm s}/ m_{3/2}c}\right) .
\end{align}
Therefore, considering any $\N=4\to 0$ super no-scale model, such as the sss one, as a ``parent" theory, one obtains automatically a chain of ``descendant"  models realizing the $\N=2\to 0$ or $\N=1\to 0$ breaking, with exponentially suppressed vacuum energy at 1-loop.


\subsection{\bm Threshold corrections without decompactification problem}
\label{decompac pb}

As shown in Ref. \cite{N=0thresh}, the gauge coupling threshold corrections of the $\Z_2^{\rm free}$  and $\Z^{\rm free}_2\times \Z_2$ descendant theories derived from no-scale models realizing the $\N=4\to 0$ breaking of supersymmetry turn out to have a universal form, free of decompactification problem. In the following, we present the running gauge coupling associated to a gauge group factor $G_\alpha$, in the $\Z^{\rm free}_2\times \Z_2$ case. At low supersymmetry breaking scale $m_{3/2}$, it is expressed in terms of moduli-dependent masses of order $m_{3/2}$ that encode the dominant contributions arising from five conformal blocks, which naturally appear in the left-moving piece of the partition function,
\be
Z_{4,0}^{(\rm F)}\!\big[{}^{a;\,  H_1,\, H_2} _{b;\; G_1,\; G_2}  \big]\, S_L\big[{}^{a;\, h} _{b;\;  g} \big] \!= {1\over 2}\sum_{a,b} (-1)^{a+b+ab} \,{\theta\big[{}^a_b\big] \over \eta}\, 
{\theta\big[{}^{a+H_2}_{b+\, G_2}\big] \over \eta}\,{\theta\big[{}^{a+H_1}_{b\, +G_1}\big] \over \eta}\,
{\theta\big[{}^{a-H_1-H_2}_{b\, -G_1-G_2}\big] \over \eta} \, (-1)^{ga+hb+hg}\, ,
\ee
associated to the 8 twisted  worldsheet fermions. 
In our conventions, $H_1,G_1\in\{0,1\}$ refer to the freely acting twists of  $\Z_2^{\rm free}$, while $H_2,G_2\in\{0,1\}$ are those of the second~$\Z_2$.

The five dominant sectors are denoted as $B$, $C$, $D$ and $I\in\{2,3\}$,  and their mass threshold scales are the following \cite{N=0thresh},  when no Wilson line deformations are switched on  : 
\begin{align}
\label{scales}
M^2_B={M^2_{\rm s}\over |\theta_2(U_1)|^4\, \Im T_1\, \Im U_1}\, , \; &M^2_C={M^2_{\rm s}\over |\theta_4(U_1)|^4\, \Im T_1\, \Im U_1}\, , \; 
M^2_D={M^2_{\rm s}\over |\theta_3(U_1)|^4\, \Im T_1\, \Im U_1}\, ,\nonumber \tesp\\
&M^2_I={M^2_{\rm s} \over 16\big\abs\eta(T_I)\abs^4 \, \big\abs\eta(U_I)\abs^4\, \Im T_I \, \Im U_I}, \quad I\in\{2,3\}\, .
\end{align}

$\bullet$  In the  conformal block $B$, the supersymmetry breaking takes place, $(h,g)\ne(0,0)$, while the $\Z_2^{\rm free}\times \Z_2$ twists are trivial, $(H_1, G_1)=(H_2, G_2)=(0,0)$. It realizes the $\N=4 \to 0$ spontaneous breaking.

$\bullet$  The conformal block $C$, with $(H_1,G_1)\ne(0,0)$ and  $(h,g)=(H_2,G_2)=(0,0)$, preserves an $\N_C=2$ supersymmetry.

$\bullet$  The conformal block $D$, with $(h,g)=(H_1,G_1)\ne(0,0)$ and $(H_2,G_2)=(0,0)$, preserves an $\N_D=2$ supersymmetry.

In the above three sectors, the $1^{\rm st}$ 2-torus is untwisted, $(H_2,G_2)=(0,0)$,  and its shifted lattice  $\Gamma_{2,2}\big[{}^{\, h,\, H_1}_{\, g,\;G_1}\big]\!_{(T_1,U_1)}$ is coupled non trivially to $Z^{(F)}_{4,0}\!\big[{}^{a;\,  H_1,\, 0} _{b;\; G_1,\; 0}  \big]$ \via the phase $S_L\big[{}^{a;\, h} _{b;\;  g} \big]$. The mass scales $M_B,M_C,M_D$ arise from the towers of KK states along the $1^{\rm st}$ 2-torus. In the blocks $C$ and $D$, where  $(H_1,G_1)\neq (0,0)$, the $2^{\rm nd}$ and $3^{\rm rd}$ 2-tori are twisted but the $1^{\rm st}$ one is shifted. Thus, there are  no massless  twisted states arising from the blocks $C$ and $D$ (no fixed points to localize them).

$\bullet$  In the remaining relevant conformal blocks $I\in\{2,3\}$, the $1^{\rm st}$ 2-torus is twisted, $(H_2,G_2)\ne (0,0)$. The $2^{\rm nd}$ 2-torus in untwisted for $I=2$, where $(H_1,G_1)=(0,0)$, while  the $3^{\rm nd}$ one is untwisted for $I=3$, where $(H_1,G_1)=(H_2,G_2)$. These blocks preserve distinct $\N_I=2$ supersymmetries. In Eq. (\ref{scales}), 
the expressions of the threshold mass scales $M_I$'s  are valid when the generator of the $2^{\rm nd}$ $\Z_2$ and its product with the generator of $\Z_2^{\rm free}$ have fixed points, namely when both $\Gamma_{2,2(T_I,U_I)}$ lattices are unshifted. 

All other conformal blocks give either vanishing contributions, like the $\N=4$ block $A$, $(h,g)=(H_1,G_1)=(H_2,G_2)=(0,0)$, or the $\N=1$ ones, which have $\big\abs{}^{H_1\;\,H_2}_{G_1\; \,G_2}\big\abs\neq 0$. Or, their contributions are  exponentially suppressed, as is the case for the  blocks $E$ and $F$, which have $(H_2,G_2)=(0,0)$ and $\big\abs{}^{h\;\,H_1}_{g\; \,G_1}\big\abs\neq0$, and realize $\N_{C}=2\to 0$ and $\N_{D}=2\to 0$ spontaneously  broken phases.  

Absorbing in a ``renormalized string coupling'' the universal contribution to the gauge coupling \cite{universality}, 
\begin{align}
{16\, \pi^2\over g_{\rm renor}^2}&={16\, \pi^2\over g_{\rm s}^2}-{1\over 2}Y(T_2,U_2)-{1\over 2}Y(T_3,U_3)\, , \phantom{\underset{\underset{\cdot}{\cdot}}{\abs}}\nonumber \\
 \with \quad  Y(T,U)&={1 \over 12}\int_{\cal F}{d^2\tau\over \tau_2}\,
\Gamma_{2,2}{}_{(T,U)} \left[ \Big(\bar E_2-{3\over \pi\tau_2}\Big){\bar E_4 \bar E_6\over \bar\eta^{24}}-\bar \jmath+1008 \right] , 
\end{align}
where $E_{2,4,6}=1+\O(q)$ are the holomorphic Eisenstein series of modular weights 2,4,6 and  
$j={1/q}+744+\O(q)$ is holomorphic and modular invariant, 
 the final result for the running gauge coupling $g_\alpha(Q)$ at energy scale $Q$ is \cite{N=0thresh},
\begin{align}
\label{thfinal}
{16\, \pi^2\over g_\alpha^2(Q)} = k^{\alpha}{16\, \pi^2\over g_{\rm renor}^2} 
&-{1\over 4}b^\alpha_{B}\log\!\left({Q^2\over Q^2+M^2_B}\right)
\! -{1\over 4}b^\alpha_{C}\log\!\left({Q^2\over Q^2+M^2_C}\right)
\! -{1\over 4}b^\alpha_{D}\log\!\left( {Q^2\over Q^2+M^2_D}\right)\phantom{\underset{\underset{\cdot}{\cdot}}{\abs}}\nonumber\tesp \\
 & -{1\over 2} b^\alpha_{2} \log\!\left({Q^2\over M^2_{2}} \right) 
 \!-{1\over 2}b^\alpha_{3}\log\!\left( {Q^2\over M^2_{3}}\right) +\O\!\left({m^2_{3/2}\over M^2_{\rm s}}\right).
\end{align}
It  only depends  on the Kac-Moody level $k^\alpha$ of the gauge group factor $G_\alpha$ and on 5 model-dependent $\beta$-function coefficients $b^\alpha_{B,C,D}$ and  $b^\alpha_{2,3}$.  The terms in the $1^{\rm st}$  line are associated to the $\N=0$, $\N_C=2$ and $\N_D=2$ spectra, which arise respectively in the conformal blocks $B$, $C$ and $D$, while those in the $2^{\rm nd}$ line arise from  the $\N_I=2$ spectra, $I\in\{2,3\}$. Note  that in the $1^{\rm st}$  line of Eq. (\ref{thfinal}), we have shifted $M_{B,C,D}^2\to Q^2+M_{B,C,D}^2$, in order to extend the validity of the result to values of $Q$ above the threshold scales $M_{B,C,D}$ at which  the  conformal blocks $B$, $C$ or $D$ decouple. Therefore, $Q$ is allowed to be as large as the lowest mass, which is of order  $cM_{\rm s}$, of the massive states we have neglected the exponentially suppressed contributions. At low energy, \ie $Q$ lower than the three scales $M_{B},M_C,M_D$, the r.h.s. of Eq. (\ref{thfinal}) behaves as $-{1\over4}(b^\alpha_B+b^\alpha_C+b^\alpha_D)\log \Im T_1+\O(1)$ when $\Im T_1$ is large and $U_1=\O(i)$. No volume term $\O(\Im T_1)$ being present, the models evade the decompactification problem. 

As already stated in the previous subsection, up to exponentially suppressed terms,  the 1-loop effective potentials in the $\Z_2^{\rm free}\times \Z_2$ models we consider here come only from the conformal block $B$ where $\N=4$ supersymmetry is spontaneously broken to $\N=0$,
\begin{align}
\V_{\mbox{\scriptsize 1-loop}}\Big\abs_{\N=2\to 1 \to 0}&={1\over 4}\, \V_{\mbox{\scriptsize 1-loop}}\Big\abs_{\N=4 \to 0}+\O\!\left(M_{\rm s}^4\, e^{-cM_{\rm s}/ m_{3/2}}\right)\phantom{\underset{\hat \abs}{\Big\abs}} \nonumber\\
&= {1\over 4}\xi(n_{\rm F}-n_{\rm B})\, m_{3/2}^4+\O\!\left(M_{\rm s}^4\, e^{-cM_{\rm s}/ m_{3/2}}\right).
\end{align}
In this expression, $n_{\rm F}-n_{\rm B}$ 
is the number of massless fermions minus the number of massless bosons in the ``parent"  $\N=4\to 0$ theory.  Actually, ${1\over 4}(n_{\rm F}-n_{\rm B})$ turns out to be the same quantity in the final $\N=1\to 0$  ``descendant'' model. This is a consequence of the
underlying  ``non-aligned" $\N_C=2$,  $\N_D=2$ and  $\N_I=2$, $I\in\{2,3\}$, supersymmetries.  Thus, when the initial $\N=4\to 0$ model is super no-scale, we have $n_{\rm F}-n_{\rm B}=0$, which guaranties the $ \Z_2^{\rm free}$ and $ \Z_2^{\rm free}\times \Z_2$ descendant orbifold theories to be  super no-scale models as well.


\subsection{\bm $T^2\times T^2\times T^2$-moduli and Wilson lines deformations}
{\label{N=2,1Deformations}}
Starting from an $\N=4\to 0$  no-scale model, the  moduli space  that survive $\Z_2$ or $\Z_2\times \Z_2$ orbifold actions in the   ``descendant'' models is reduced. This follows from the fact that several deformations are  frozen to some discrete values, in order to respect the factorization of the internal 6-torus as $T^2\times T^4$ or $T^2\times T^2\times T^2$. For instance, in the sss-model, the scalars $Y_i$ in Eq. (\ref{Yi}) are fixed to 0. However, new moduli fields arise generically from the massless scalars of the twisted sectors. Therefore, the stability and quantum flatness condition of the $\N=2\to 0$ and $\N=1\to 0$ no-scale models must be reconsidered. 

An exception however exists, for the models arising from $\N=4\to 0$ no-scale theories, on which a $\Z_2^{\rm free}$ or $\Z^{\rm free}_2\times \Z_2$ orbifold action is implemented, as described in Subsect. \ref{chain}. In this case, modulo the constraint of the $\Gamma_{6,6}$ lattice factorization, the structure of the deformed effective potential  is  as in  Eq. (\ref{Vtot}), up to the multiplicative factor ${1\over 2}$ or ${1\over 4}$, and fully arises  from the untwisted sector. Due to the free action of $\Z_2^{\rm free}$, the $1^{\rm st}$ 2-torus is not fixed under any orbifold group element, so that no twisted massless states and thus no new moduli sensitive to the stringy Scherk-Schwarz mechanism is introduced. On the contrary, twisted massless states are allowed in the conformal blocks where the $2^{\rm nd}$ or $3^{\rm rd}$ 2-tori are fixed. However, being $\N_2=2$ or $\N_3=2$ supersymmetric at tree level, new moduli deformations exist, but remain exactly flat directions at 1-loop and therefore do not show up in the effective potential at this order. Thus, in the study of the quantum stability of the $\N=4 \to 2\to 0$ or $\N=2\to 1\to 0$ models obtained by $\Z^{\rm free}$ or $\Z^{\rm free}_2\times \Z_2$ orbifold actions, only the $\beta$-function coefficients of the ``parent" $\N=4\to 0$ theory are relevant. The resolution of an instability in a chain of $\N=4,2,1\to 0$ no-scale models is thus universal, in the sense that it is independent of the specific spectra of the ``descendant" theories. 


 \section{Conclusion}
 \label{cl}
 
 In this work, we focus on no-scale string models \cite{noscale} where the spontaneous $\N=4,2,1\to 0$ breaking of supersymmetry  is implemented  at the perturbative level by geometrical fluxes. This setup realizes  a  ``coordinate-dependent string compactification" \cite{SSstring, Kounnas-Rostand}, in the spirit of the Scherk-Schwarz mechanism introduced in supergravity \cite{SS}. The gravitino mass scale $m_{3/2}$ is  related to the inverse volume of the compact space involved in the supersymmetry breaking.
Even thought  supersymmetry is broken, the classical effective potential is  positive semi-definite, $\V_{\rm tree} \ge 0$ \cite{noscale}, while the   supersymmetry breaking scale  $\langle m_{3/2}\rangle$ is undetermined  by the flatness condition. 
 
At the quantum level, the 1-loop effective potential receives non-trivial corrections. The latter are however under  control, at least in the regime of low supersymmetry breaking scale, $m_{3/2}<cM_{\rm s}$, in which case one has
\be
\V_{\mbox{\scriptsize 1-loop}} =\xi(n_{\rm F}-n_{\rm B})\, m_{3/2}^4+\O\!\left(M_{\rm s}^4\, e^{-c M_{\rm s}/ m_{3/2}}\right) .
\ee
The above formula arises from the contributions of the light KK towers of states associated to the large internal space, and remains valid in the string context we consider   {\em even when the no-scale models realize the $\N=2\to 0$ or $\N=1\to 0$ breaking}. These facts lead us to consider the situation where the numbers of massless fermionic and bosonic degrees of freedom are equal, $n_F=n_B$ \cite{ADM,planck2015}.  In this case, $\V_{\mbox{\scriptsize 1-loop}}$ vanishes modulo exponentially suppressed terms and we refer to these theories as ``super no-scale string models". At the 1-loop level, they satisfy the flatness condition, as well as the absence of dilaton  and no-scale modulus tadpoles. 

Simple examples of  ``super no-scale models"  are constructed in the framework of the heterotic string compactified on $T^2\times T^2\times T^2$. They realize the $\N=4\to 0$ spontaneous breaking \via a stringy Scherk-Schwarz mechanism along the $1^{\rm st}$ internal 2-torus and  their right-moving gauge symmetry is either 
\be
(a) \;\;\G=U(1)^2\times SU(2)^4\times  SO(16)^2\quad \mbox{or} \quad (b) \;\;U(1)^3 \times SU(2)\times SU(3)\times  SO(16)^2\, .
\ee
In both examples, if $\V_{\mbox{\scriptsize 1-loop}}$ is exponentially suppressed when $m_{3/2}<M_{\rm s}$, it is not suppressed when $m_{3/2}=\O(M_{\rm s})$. However, no  Hagedorn-like instability takes place in this regime  \cite{Hage, Kounnas-Rostand}, which means that no state becomes tachyonic at any point of the $(T_1,U_1)$-moduli space. Moreover, in the regime where $m_{3/2}>M_{\rm s}$, the model is more naturally interpreted as an explicitly non-supersymmetric theory, rather than a no-scale model. Altogether,  $\V_{\mbox{\scriptsize 1-loop}}$ turns out to be positive and increases monotonically with $m_{3/2}$. Therefore, in a cosmological context, the dynamics drives naturally these models to the super no-scale regime, where the supersymmetry breaking scale is small.

We also examine the local stability of the model with gauge symmetry ${\cal G}=U(1)^2\times SU(2)^4\times SO(16)^2$, under small moduli perturbations of the internal $\Gamma_{6,6+16}$ lattice. The analysis actually applies to all no-scale string models realizing an $\N=4\to 0$ breaking \via stringy Scherk-Schwarz mechanism  \cite{SSstring, Kounnas-Rostand} along a large $1^{\rm st}$ internal 2-torus, wether they are super no-scale, \ie with $n_{\rm F}=n_{\rm B}$, or not. The rank of the gauge group being always $6+16$, we find the following three possible behaviors of the moduli $Y_{IJ}$, $I\in \{1,\dots, 6\}$, $J\in\{1,\dots, 6+16\}$~: 

$\bullet$  For $J$ associated to a Cartan generator of an asymptotically free gauge group factor  $G_\alpha$ ($b_\alpha<0$), the $Y_{IJ}$'s acquire 1-loop masses of order $m_{3/2}$,  and are  therefore stabilized at the origin, $Y_{IJ}=0$. 

$\bullet$  For $J$ corresponding to a Cartan generator of a non-asymptotically free gauge group factor ($b_\alpha>0$), the $Y_{IJ}$'s aquiere negative squared masses, which leads instabilities. They condense and break $G_\alpha$ to subgroups with non-negative $\beta$-function coefficients. 

$\bullet$  The last $Y_{IJ}$'s, associated to gauge group factors with $b_\alpha=0$, remain massless. 

Thus, in the examples we considered,  the $SO(16)\times SO(16)^{\prime}$ Wilson lines  yield  a destabilization of the initial background. However, we stress  that in all super no-scale models, the quantum instabilities are harmless when the theories are considered at finite temperature $T$, provided that $T>m_{3/2}$. This follows from the fact that finite temperature induces effective mass terms proportional to $T^2 (Y_{IJ})^2$, which screen all tachyonic contributions  $-m^2_{3/2} (Y_{IJ})^2$. Therefore,  in the framework of string cosmology  at finite temperature \cite{CosmologicalTerm}, a phase transition happens  when $T$ approaches $m_{3/2}$ from above, which drives  the initial model to a new phase without non-asymptotically free gauge group factors.

A particular class of super no-scale models, which realize the spontaneous $\N=2\to 0$ or $\N=1\to 0$ breaking of supersymmetry, can be  constructed easily. They are built from parent $\N=4\to 0$ super no-scale models, on which a $\Z^{\rm free}_2$ or $\Z^{\rm free}_2 \times \Z_2$ orbifold action is implemented. The fact that the $\Z^{\rm free}_2$ group is freely acting ensures that the partial $\N=4\to 2$ breaking is {\it spontaneous}, which yields important consequences \cite{N=0thresh}.  First,  the 1-loop effective potential in the descendant models is simply  $1\over 2$ or $1\over 4$ of that of the parent theory. Second, the threshold corrections to the gauge couplings are {\em not proportional} to the volume of the large internal submanifold involved in the stringy Scherk-Schwarz mechanism. This fact guaranties the validity of the string perturbative expansion, \ie   solves the so-called ``decompactification problem".  In the descendent theories, the space of untwisted moduli, which are those appearing in the effective potential, is reduced, as follows from the factorization of the internal space required by the orbifold action. 

To conclude, we mention that it would be very interesting to study in  super no-scale models the higher order corrections in string coupling to the effective potential. This would allow to  see wether insisting on the flatness condition would yield additional  restrictions on the models. One can also construct super no-scale theories by implementing $\Z_2$ or $\Z_2\times \Z_2$ orbifold  actions on  $\N=4\to 0$ no-scale models, where  each $\Z_2$ admits fixed points \cite{ADM}. In this case, our analysis of the moduli deformations must be completed, since the effective potential {\em does} depend on twisted moduli sensitive to the final breaking of supersymmetry to $\N=0$. However, if these models are compatible with the physical requirement of possessing chiral spectra, the decompactification problem has to be readdressed.  


\section*{Acknowledgement}
 
We are grateful to  S. Abel, C. Angelantonj, C. Bachas, G. Dall'Agata, A. Faraggi, I. Florakis, I. Antoniadis and  J. Rizos  for fruitful discussions. 
The work of C.K. is partially supported by his  Gay Lussac-Humboldt Research Award 2014, in the  Ludwig Maximilians University and Max-Planck-Institute for Physics. H.P. would like to thank the Laboratoire de Physique Th\'eorique of Ecole Normale Sup\'erieure and the C.E.R.N. Theoretical Physics Department for hospitality. 



\end{document}